\documentclass[11pt]{article}
\usepackage{graphicx}
\graphicspath{{./figures/}}
\usepackage{appendix}
\usepackage{latexsym,amsmath,amsfonts,amssymb,booktabs}
\usepackage[font=small]{caption}
\usepackage{slashed,upgreek,amscd,cancel,tensor,color}
\usepackage{adjustbox}
\usepackage[numbers,compress,square]{natbib}
\usepackage{epsfig,latexsym}
\usepackage{amsmath}
\usepackage[pdfencoding=auto]{hyperref} 
\usepackage{url}
\numberwithin{equation}{section}
\usepackage{doi}
\usepackage{subcaption}
\definecolor{MyBlue}{rgb}{0.15,0.15,0.70}

\hypersetup{
colorlinks=true,
citecolor=MyBlue,
linkcolor=MyBlue,
urlcolor=MyBlue
}

\input epsf
\usepackage{amssymb}
\usepackage{amsmath}
\usepackage{amsfonts}
\usepackage{upgreek}
\usepackage{latexsym}
\usepackage{stfloats}
\usepackage{afterpage}

\newcommand{\be}{\begin{equation}}
\newcommand{\ee}{\end{equation}}
\newcommand{\beq}{\begin{equation}}
\newcommand{\eeq}{\end{equation}}
\newcommand{\bea}{\begin{eqnarray}}
\newcommand{\eea}{\end{eqnarray}}

\usepackage[stable]{footmisc}

\def\d{\delta}

\def\dkmu2{\delta K_{\mu \nu}\delta K^{\mu \nu}}
\def\pmu2{  \phi_{\mu \nu}\phi^{\mu \nu}}

\newcommand{\epsilonosc}{\epsilon_{\rm osc}}
\newcommand{\ldv}{L_{\Delta v}}
\newcommand{\ldvd}{\dot L_{\Delta v}}

\renewcommand\[{\left[}

\newcommand\ees{\end{eqnarray}}
\newcommand\bees{\begin{eqnarray}}

\newcommand\alphaH{\alpha_{\text{H}}}

\newcommand\alphaV{\alpha_{\text{V}}}
\newcommand\alphaB{\alpha_{\text{B}}}
\newcommand\alphaM{\alpha_{\text{M}}}
\newcommand\alphaK{\alpha_{\text{K}}}
\newcommand\alphaT{\alpha_{\text{T}}}

\newcommand{\unitsk}{ \, h \, \text{Mpc}^{-1}}
\newcommand{\unitsr}{ \, h^{-1} \text{Mpc}}
\newcommand{\kvec}{\vec{k}}
\newcommand{\qvec}{\vec{q}}

\newcommand{\eqn}[1]{eq.~(\ref{#1})}

 \newcommand{\momspmeas}[1]{\frac{d^3 #1}{(2 \pi)^3}}
 \newcommand{\half}{\frac{1}{2}}

\newcommand{\xvec}{\vec{x}}

\newcommand{\mpl}{M_{\rm Pl}}

\newcommand{\secref}[1]{Sec.~\ref{#1}}

\newcommand{\appref}[1]{App.~\ref{#1}}
\newcommand{\figref}[1]{Fig.~\ref{#1}}

\newcommand{\rvec}{\vec{r}}

\newcommand{\diffsq}{| \vec{k} - \vec{q}|^2}

\newcommand{\esl}{\epsilon_{s<}}
\newcommand{\edl}{\epsilon_{\delta <}}
\newcommand{\esg}{\epsilon_{s>}}
\newcommand{\om}{\Omega_{\rm m}}
\newcommand{\ttt}{\tilde t}
\newcommand{\bun}{\beta_1}
\newcommand{\bdeux}{\beta_2}
\newcommand{\btrois}{\beta_3}
\newcommand{\eqns}[2]{eqs.~(\ref{#1} - \ref{#2})}
\newcommand{\barrhom}{\bar \rho }
\newcommand{\ellbao}{\ell_{\rm BAO}}

\begin{document}
\vspace{0.5cm}

\vspace{0.5cm}

\begin{center}
\Large{\textbf{Violation of the consistency relations \\   for large-scale structure with dark energy}} \\[1cm]

\large{Matthew Lewandowski$^{1,2}$}
\\[0.5cm]

\small{
\textit{$^{1}$ Department of Physics and Astronomy, \\ Northwestern University, Evanston, IL 60208, USA }}

\vspace{.2cm}

\small{
\textit{$^2$ Institut de physique th\' eorique, Universit\'e  Paris Saclay, \\ 
CEA, CNRS, 91191 Gif-sur-Yvette, France  }}

\vspace{.2cm}

\vspace{0.5cm}
\today

\end{center}

\vspace{2cm}

\begin{abstract}
We study infrared effects in perturbation theory for large-scale structure coupled to the effective field theory of dark energy, focusing on, in particular, Degenerate Higher-Order Scalar-Tensor (DHOST) theories.  In the subhorizon, Newtonian limit, DHOST theories introduce an extra large-scale velocity $v^i_\pi$ which is in general different from the matter velocity $v^i$.  Contrary to the case in Horndeski theories, the presence of this extra large-scale velocity means that one cannot eliminate the long-wavelength effects of both $v^i$ and $v^i_\pi$ with a single coordinate transformation, and thus the standard $\Lambda$CDM consistency relations for large-scale structure are violated by terms proportional to the relative velocity $v^i - v^i_\pi$.  We show, however, that in non-linear quantities this violation is determined by the linear equations and the symmetries of the fluid system.  We find that the size of the baryon acoustic oscillations in the squeezed limit of the bispectrum is modified, that the bias expansion contains extra terms which contribute to the squeezed limit of the galaxy bispectrum, that infrared modes in the one-loop power spectrum no longer cancel, and that the equal-time double soft limit of the tree-level trispectrum is non-vanishing.  {We also show how one can resum the effects of \emph{one} of the large-scale velocities on the power spectrum, and how to include the relative velocity perturbatively in the resummation.  This result is also applicable to other theories with a large-scale relative velocity, such as dark matter and baryons.}  Many of our computations are also relevant for perturbation theory in $\Lambda$CDM with exact time dependence. 

\end{abstract}

\newpage

\tableofcontents

\vspace{.5cm}
\newpage

%
%


\section{Introduction} \label{introsec}

Large-scale structure (LSS) surveys, which measure the positions and redshifts of galaxies to provide a three-dimensional map of the galaxy density field, could very well be the next leading sources of cosmological information.  These surveys provide a rich picture of our cosmology.  What we see is the final state of the evolution of initial conditions, set by inflation, through the history of the universe: from the radiation dominated early stages to the matter and cosmological-constant dominated current era.  In that time, perturbations in the curvature of space-time get imprinted on the densities of the particle species present in the universe.  The initial seeds of curvature fluctuations imprint perturbations on, among other species, the cold dark-matter (CDM) density field, whose evolution quickly becomes dominated by gravity.  As the dark matter clumps, galaxies tend to form in the regions of high density, so by measuring the properties of the distribution of galaxies, we can learn about both the initial conditions of the universe and their subsequent evolution.  All of this has occurred in the background of an expanding universe, which has only recently changed from being matter dominated to being dominated by the cosmological constant $\Lambda$.  This standard picture of the history of the universe is called $\Lambda$CDM.

Precisely connecting the galaxy-density map to fundamental cosmological information is no small theoretical challenge, though.  One must understand, at the percent level, non-linear dark-matter clustering, the effects of sub-dominant species like baryons and massive neutrinos, the non-linear evolution of the baryon acoustic oscillations (BAO), redshift-space distortions, and galaxy biasing.  This tremendous task has been undertaken with standard techniques (see for example \cite{Alam:2016hwk, Beutler:2016ixs,Beutler:2016arn} and references therein) but as improved surveys provide unprecedented, percent-level data, a robust and theoretically controllable framework must be used (see for example \cite{DAmico:2019fhj,Ivanov:2019pdj,Colas:2019ret} and references therein).  The latter framework, called the effective field theory of large-scale structure (EFT of LSS) \cite{Baumann:2010tm,Carrasco:2012cv}, was developed to provide a well-defined perturbative expansion of clustering statistics to access short scale, mildly non-linear modes in galaxy surveys.  Only after all of this is understood can one can begin to probe new physics such as non-Gaussianity and dark energy.  For the EFT of LSS treatment of the above issues, see for example \cite{Baumann:2010tm,Carrasco:2012cv, Porto:2013qua, Carrasco:2013sva, Mercolli:2013bsa, Lewandowski:2014rca, Senatore:2017hyk, Lewandowski:2016yce, Cusin:2017wjg, Senatore:2014eva, Senatore:2014vja, Lewandowski:2015ziq, Foreman:2015uva}.  

Complementing precision perturbation theory for LSS, we have another important tool for understanding cosmological perturbations, namely the consistency relations for LSS in $\Lambda$CDM \cite{Peloso:2013zw, Kehagias:2013yd, Creminelli:2013mca}, which relate soft (i.e. small momenta) limits of $(n+1)$-point correlation functions to $n$- or lower-point correlation functions (see \cite{Creminelli:2013poa} for an extension to multiple soft limits and redshift space, and \cite{Esposito:2019jkb} for a verification of the consistency relations in $N$-body simulations).  This kind of consistency relation was first pointed out in the context of single-field inflation \cite{Maldacena:2002vr}.  These relations are generally non-perturbative and can be derived using a coordinate transformation and shift of the fields \cite{Maldacena:2002vr,Creminelli:2004yq} based on the equivalence principle \cite{Creminelli:2013mca}.  The form of the transformation in the non-relativistic limit is that of a Galilean transformation, i.e. a boost into a frame with a given velocity.  By boosting to the frame comoving with the dark matter, one eliminates the effect of the large-scale velocity (i.e. gradient of the metric): this is the basis of the consistency relations.  Crucial to this construction is that the velocities of all species can be eliminated in this way, i.e. that there is a dominant adiabatic mode \cite{Weinberg:2003sw, Horn:2014rta}.  

In this paper, we study the effects of dark energy on the consistency relations for LSS, naturally extending the work of \cite{Crisostomi:2019vhj} to the BAO, bias expansion, one-loop matter power spectrum, and tree-level trispectrum.  The presence of dark energy (a light scalar field $\pi$), can violate the consistency relations by introducing a second large-scale velocity, $v_{\pi}^i \propto \partial_i \pi$, in addition to the matter velocity $v^i$.  As long as the relative velocity $v^i - v^i_\pi \neq 0$, one cannot eliminate both long-wavelength velocity fields with a single Galilean transformation.  Thus we expect (and indeed confirm), that the consistency relations will be violated by terms proportional to the relative velocity.  To be precise, by violation of the consistency relations, we simply mean that the standard $\Lambda$CDM consistency relations are no longer true, because one cannot construct a physical adiabatic mode.

The situation here is similar to the case of multiple fluids, like CDM and baryons \cite{Peloso:2013spa, Bernardeau:2011vy, Bernardeau:2012aq, Creminelli:2013poa, Lewandowski:2014rca}, where the presence of a relative velocity also violates the consistency relations.  In that case, however, the relative velocity decays as $a^{-1}$ ($a$ is the scale factor of the metric, see \eqn{metric}), and so is negligibly small at late times, but not necessarily so at early times \cite{Tseliakhovich:2010bj}.  The difference, though, is that the dark-energy field $\pi$ is not dynamical in the Newtonian limit (it is a constraint field like the gravitational potential $\Phi$), and so there are not really two fluids (see \cite{Lewandowski:2017kes} for an example of a dynamical dark energy, clustering quintessence, which preserves the consistency relations).  What happens is that the equations of motion, written in terms of only the dark-matter overdensity, get modified by the $\pi$ field to essentially contain two different convective derivatives.  This means that one cannot generally construct a physical adiabatic mode.  Although the consistency relations are violated, we find that the leading effects in the infrared (IR) can be determined by the linear equations and the symmetries of the fluid system.  This allows us to study the leading IR effects, in particular those that violate the consistency relations, without having to solve the full equations of motion.  Because our results are determined by the symmetries in the IR, it would also be fair to call them \emph{modified} consistency relations.

In this work, we describe the dark-matter fluid by the overdensity $\delta \equiv ( \rho - \bar \rho) / \rho$, where $\rho$ is the mass density of the fluid and $\bar \rho$ is the time-dependent background value, and the velocity field $v^i$.  The fluid moves under the forces of the Newtonian gravitational potential $\Phi$, which appears in the Newtonian gauge metric (considering only scalar perturbations),
\be \label{metric}
ds^2 = - ( 1 + 2 \Phi ) dt^2 + a(t)^2 (1 - 2 \Psi ) d \vec x^2 \ . 
\ee
The fluid equations for the dark matter field are (valid in the non-relativistic, Newtonian limit $v/c \ll 1$ and $c k / H \gg 1$, where $k$ is the wavenumber),
\begin{align}
\begin{split} \label{fluideqs}
& \dot \delta + a^{-1} \partial_i \left( ( 1 + \delta) v^i \right) = 0  \ ,  \\
& \dot v^i + H v^i +a^{-1} v^j \partial_j v^i + a^{-1} \partial_i \Phi = - (a \rho)^{-1} \partial_j \tau^{ij}  \ ,
\end{split}
\end{align}
where $H = \dot a / a$ is the Hubble rate (a dot denotes a derivative with respect to $t$), and $\tau^{ij}$ is the stress tensor generated by short-scale non-linearities (i.e. counterterms) \cite{Baumann:2010tm, Carrasco:2012cv}.  In this work we set the speed of light to unity $c = 1$, so that all velocities are measured with respect to $c$.  To complete the fluid equations \eqn{fluideqs}, one needs to relate the gravitational potential $\Phi$ to the dark-matter density.  In $\Lambda$CDM, this comes through the Poisson equation,
\be
 a^{-2} \partial^2 \Phi = \frac{3}{2} H^2 \om \delta \ , 
\ee
where $\om = \bar \rho / ( 3 \mpl^2 H^2)$ is the time-dependent matter fraction, and $\mpl$ is the Planck mass.  However, in theories with dark energy, this relationship is modified.

One way to study possible deviations from general relativity on cosmological scales is to consider theories which break time diffeomorphisms.  In order to avoid unstable Ostrogradsky modes \cite{Woodard:2015zca}, it is common to focus on theories which propagate a single scalar degree of freedom.  These include Horndeski theories \cite{Horndeski:1974wa, Deffayet:2011gz}, where the equations of motion are at most second order in the metric and scalar field.  This class of theories can be further extended to include equations of motion with more than two derivatives, but which still propagate a single scalar mode.  This class is known as Degenerate Higher-Order Scalar-Tensor (DHOST) theories \cite{Langlois:2015cwa, Crisostomi:2016czh, BenAchour:2016fzp}, a subset of which are the theories beyond Horndeski, such as Gleyzes-Langlois-Piazza-Vernizzi (GLPV) theories \cite{Gleyzes:2014dya} (see also \cite{Zumalacarregui:2013pma} for examples beyond Horndeski).  In particular, in this work, we consider non-linear DHOST theories within the EFT of dark energy  \cite{Dima:2017pwp,Crisostomi:2019yfo} (see \cite{Langlois:2017mxy} for the linear theory).  Using the ADM decomposition of the metric $ds^2 = - N^2 dt^2 + h_{ij} (dx^i + N^i dt) (dx^j + N^j dt)$, and choosing time to coincide with the uniform scalar-field hypersurfaces, this is given by
\be
\begin{split}
\label{EFTaction}
&S_{\rm EFT} = \int  d^4 x \sqrt{h}  \frac{M^2}{2} \big[ -(1+\delta N)\delta {\cal K}_2 +c_{\rm T}^2 {}^{(3)}\!R   + H^2 \alphaK \delta N^2 + 4 H \alphaB \delta K \delta N    \\
 & \hspace{.7in} +(1+\alphaH) {}^{(3)}\!R \delta N +4 \beta_1 \delta K V + \beta_2 V^2 + {\beta_3} a_i a^i
+ \alphaV \delta N \delta {\cal K}_2 \big]  \ , 
\end{split}
\ee
where we have only included operators with the highest number of spatial derivatives, which are the dominant terms in the Newtonian limit.  Here $\delta N \equiv N-1$, $\d K_i^j \equiv K_i^j- H \delta_i^j$  is the perturbation of the extrinsic curvature of the time hypersurfaces, $\delta K$ its trace, and ${}^{(3)}\!R$ is the 3D Ricci scalar of these hypersurfaces. Moreover, $\delta {\cal K}_2 \equiv \delta K^2 - \delta K_{i}^j \delta K^{i}_{j}$, $V \equiv (\dot N - N^i \partial_i N)/N${,} and $a_i \equiv \partial_i N/N$.   

In the action \eqn{EFTaction}, Horndeski theories have $\alphaH = \beta_1=\beta_2=\beta_3=0$, in which case there are four free time-dependent functions: $\alphaK$, $\alphaB$, $c_{\rm T}^2$ and $\alphaM \equiv d \log M^2/d \log a$  \cite{Bellini:2014fua}.  In the Newtonian limit, the function $\alphaK$ only appears in the expression for the speed of sound of the scalar field.  Beyond Horndeski theories have $\alphaH \neq 0$, and DHOST theories have $\beta_1 \neq 0$, while the functions $\beta_2$ and $\beta_3$ are given in terms of $\beta_1$ by the degeneracy conditions \cite{Langlois:2017mxy} $\bdeux=-6\bun^2$,  $ \btrois=-2\bun\left[2(1+\alphaH)+\bun c_{\rm T}^2 \right]$ ,
which we will always impose.\footnote{Many of the parameters of these theories are constrained by observations.  The measurement of the relative light and gravitational-wave speeds \cite{Monitor:2017mdv} places significant constraints \cite{Creminelli:2017sry, Sakstein:2017xjx, Ezquiaga:2017ekz, Baker:2017hug}.  The breaking of Vainshtein screening inside of astrophysical sources \cite{Kobayashi:2014ida} places further constraints \cite{Crisostomi:2017lbg, Langlois:2017dyl, Dima:2017pwp} (see \cite{Saltas:2019ius} for a recent improvement).  More constraints can be placed by suppressing gravitational-wave decay into dark energy \cite{Creminelli:2018xsv, Creminelli:2019nok} (which forces $\alphaH = - 2 \beta_1$) and instabilities induced by gravitational waves \cite{Creminelli:2019kjy}.  Vainshtein screening for theories evading the above constraints was studied in \cite{Hirano:2019scf, Crisostomi:2019yfo}.  Finally, theoretical arguments using positivity have dramatically reduced the cutoff for some EFTs of scalars coupled to gravity \cite{PhysRevLett.123.251103}, for example the cubic galileon \cite{Nicolis:2008in}.  If this is the case, then constraints from LSS, which would be within the regime of validity of the EFT, may be more relevant than constraints coming from smaller scales, as in Vainshtein screening, which may be outside of the regime of validity of the EFT. }    Furthermore, for simplicity, for the rest of this work we will assume that $c_{\rm T}^2 = 1$.  For the same reason, we have left off some cubic operators in the action \eqn{EFTaction} which would generically renormalize $c_{\rm T}^2$ and push it away from unity (see for example \cite{Creminelli:2017sry}).  We also assume that matter is minimally coupled to the gravitational metric $g_{\mu \nu}$, i.e.
\be
S_{\rm m} = - \int d^3 x \, dt \, a^3 \bar \rho \, ( 1 + \delta ) \Phi
\ee
in the Newtonian limit, and that all species couple universally.

The action \eqn{EFTaction} is in the unitary gauge, so the new scalar field is contained in the metric (there is an extra scalar degree of freedom in the metric because the action breaks time diffeomorphisms).  One can then explicitly introduce the scalar mode by using the St\"uckelberg trick (essentially by performing a time diffeomorphism $t \rightarrow t + \pi ( \xvec , t)$, exactly as in the EFT of inflation \cite{Cheung:2007st}).  After this, the action \eqn{EFTaction} depends on the three fields $\Phi$, $\Psi$, and $\pi$.  After canonically normalizing the $\pi$ field, one can see that the unitarity cutoff for the $\pi$ interactions in \eqn{EFTaction}, for $\mathcal{O}(1)$ dimensionless couplings, is around $\Lambda_3 \sim (\mpl H_0^2)^{1/3} \approx 1/(1000 \text{ km})$, where $H_0$ is the Hubble rate today.  This means that the scales relevant for LSS are well within the EFT description of dark energy.

In the Newtonian limit that we consider in this paper, time derivatives of the gravitational fields $\Phi$, $\Psi$, and $\pi$, which are of order the Hubble scale $H$, are suppressed with respect to spatial derivatives, so the equations of motion become constraint equations.  By solving the constraint equations for these fields and perturbatively including the coupling to matter, we obtain a modified Poisson equation of the form
\be
a^{-2} \partial^2 \Phi = \mathcal{N} [ \delta ] \ , 
\ee
where $\mathcal{N}$ is some non-linear functional of the matter density that depends on the EFT action \eqn{EFTaction}.  This completes the system of equations for $\delta$, which we can then solve perturbatively.

The structure of this paper is as follows.  In \secref{lcdmpertsec}, we review perturbation theory in $\Lambda$CDM.  In \secref{crlcdmreview} we review the standard consistency relations in $\Lambda$CDM and also show how they can be proven in perturbation theory with exact time dependence (i.e. using the Green's function).  In \secref{dhostsec} we present the gravitational field equations for DHOST theories, discuss their leading IR properties, and solve for the leading IR contributions to the second and third order matter overdensity.  Contrary to the case in Horndeski theories, we find that the leading IR solutions in DHOST theories deviate from the standard $\Lambda$CDM solutions.  In \secref{observablessec}, we discuss how these solutions violate the consistency relations, in particular focusing on the BAO oscillations in the squeezed limit of the bispectrum, the bias expansion, the one-loop power spectrum, and the tree-level trispectrum.  {Furthermore, we discuss the IR-resummation, which is necessary to correctly describe the BAO in correlation functions, in this context.  We show that one can resum the effects of \emph{one} of the large-scale velocities, but that the effect of the relative velocity must be treated perturbatively.  Our results are also applicable to correctly performing the IR-resummation with multiple fluids, such as with dark matter and baryons.}
In \secref{conclusionssec}, we conclude.  
We have reserved the Appendix for additional details and discussion: in \appref{crreviewapp} we present the background wave argument proof of the standard consistency relations, in \appref{gfapp} we provide details of the Green's function manipulations used in the main text, in \appref{sec:coeff} we provide the relationships between the coefficients in the dark-energy action \eqn{EFTaction} and the equations of motion, {and in \appref{resumapp} we give details related to the IR-resummation with a large-scale relative velocity.}

We use Latin indices like $i, j , k , \dots$ to denote spatial components, and we do not distinguish between upper and lower spatial indices.  We denote the spatial laplacian by $\partial^2 \equiv \sum_{i=1}^3 \partial_i \partial_i$, and use the three-dimensional Levi-Civita symbol $\varepsilon^{ijk}$, normalized so that $\varepsilon^{123} =1$, and with no factors of the metric.  For numerical computations, we use the $\Lambda$CDM cosmological parameters $\Omega_{\rm m , 0} = 0.281$, $\Omega_{\rm b, 0} = 0.046$, $h = 0.697$, $\Delta \zeta^2 = 2.37 \times 10^{-9}$, and $n_s = 0.971$.

%
%
%

\section{Perturbation theory in $\Lambda$CDM} \label{lcdmpertsec}

The dark-matter fluid system is roughly linear on large scales and becomes non-linear on smaller scales.  Thus, it is common to solve the dark-matter system perturbatively.  To do that, one expands the overdensity like 
\be \label{expanddelta}
\delta = \delta^{(1)} + \delta^{(2)} + \delta^{(3)} + \dots \ , 
\ee
where $\delta^{(1)}$ solves the linear equations of motion, and $\delta^{(n)} \sim [\delta^{(1)}]^n$ are iterative non-linear corrections.  In particular, we assume the growing mode solution 
\be \label{deltasol}
\delta^{(1)}_{\kvec}(t) = \frac{D_+(t)}{D_+(t_{\rm in})} \delta_{\kvec}^{\rm in}\ , 
\ee
where $t_{\rm in}$ is the time at which we set the initial conditions, and $D_+ ( t ) $ is the fastest growing solution to
\be \label{growingmodeeq}
\ddot D_+ + 2 H \dot D_+ - \frac{3 H^2 \om}{2} D_+ = 0 \ .
\ee
With the linear solution for $\delta$ in \eqn{deltasol}, we can also write the linear solution for the velocity using \eqn{fluideqs}
\be \label{linearvel}
v^{i}_{(1)} = - a \frac{\partial_i \dot \delta^{(1)}}{\partial^2 } =  - a H f \frac{\partial_i  \delta^{(1)}}{\partial^2 } \ ,
\ee
where we have introduced the linear growth rate 
\be
f \equiv \frac{\dot D_+}{H D_+} \ . 
\ee

The equations of motion \eqn{fluideqs} can then be solved by iteratively plugging in the lower order solutions to get equations of the form
\be
\ddot \delta^{(n)}_{\kvec} ( t )  + 2 H (t )  \dot \delta^{(n)}_{\kvec} (  t)  - \frac{3 H(t)^2 \om(t)}{2}  \delta^{(n)}_{\kvec} ( t )  = S^{(n)} [ \kvec ; t ] \ , 
\ee
where $S^{(n)}  \sim [\delta^{(1)}]^n $ is the $n$-th order source term, which can be found using \eqn{fluideqs}.  The $n$-th order solution can then be obtained by applying the Green's function to the source term as
\be \label{generalgf}
\delta^{(n)}_{\kvec} ( t ) = \int_0^t d t_1  \, G ( t , t_1 ) \, S^{(n)} [ \kvec ; t_1 ]  \ , 
\ee
where the Green's function satisfies\footnote{In terms of two linearly independent solutions $D_+$ and $D_-$ of \eqn{growingmodeeq}, the retarded Green's function, which satisfies the boundary conditions $G( t_1 , t_1 ) = 0$ and  $\partial_t G ( t , t_1) |_{t = t_1 + \epsilon} = 1 $ as $\epsilon \rightarrow 0_+$,  is explicitly given by
\be \label{gfdef}
G(t, t_1 ) = W(t_1)^{-1} \left( D_- ( t ) D_+ ( t_1) - D_+ ( t ) D_- ( t_1) \right) \Theta_{\rm H} ( t - t_1)  \ ,
\ee
where $\Theta_{\rm H}$ is the Heaviside step function and the Wronskian is given by 
\be
W( t ) = D_+ (t ) \dot D_-(t) - D_-(t) \dot D_+(t) \ . 
\ee
We will often use the notation $\bar G( t , t_1) \equiv W(t_1)^{-1} \left( D_- ( t ) D_+ ( t_1) - D_+ ( t ) D_- ( t_1) \right)$ to refer to the part of the Green's function with no Heaviside function.  }
\be 
\partial^2_t G ( t , t_1) + 2 H(t)  \partial_t G ( t , t_1 ) - \frac{3 H(t)^2 \om ( t )}{2} G(t,t_1 ) = \delta_D ( t - t_1)  \ .
\ee
In general, we can also write the solution \eqn{generalgf} in powers of $\delta^{(1)}$ as\footnote{In this paper, we use the following notation 
\be
\int_{\kvec_1 , \dots , \kvec_n} \equiv \int \frac{d^3 k_1 }{(2 \pi)^3} \cdots \frac{d^3 k_n}{(2 \pi)^3 } \ ,  \quad \int_{\kvec_1 , \dots , \kvec_n}^{\kvec} \equiv \int_{\kvec_1 , \dots , \kvec_n} ( 2 \pi)^3 \delta_D ( \kvec - \sum_{i = 1}^n \kvec_i )  \ , 
\ee
where $\delta_D$ is the Dirac delta function, and our Fourier conventions are
\be
f ( \xvec , t ) = \int_{\kvec} f_{\kvec} ( t ) \,  e^{i \kvec \cdot \xvec}  \ . 
\ee
For a three-dimensional vector $\kvec$, we write $k \equiv | \kvec |$ for the magnitude, and $\hat k \equiv \kvec / k$ for the unit vector parallel to $\kvec$.  
}
\be
\delta^{(n)}_{\kvec }( t ) = \int_{\kvec_1, \dots , \kvec_n}^{\kvec} F_n ( \kvec_1 , \dots , \kvec_n ; t ) \delta^{(1)}_{\kvec_1 } ( t ) \cdots \delta^{(1)}_{\kvec_n} ( t ) \ ,
\ee
which defines the perturbation theory kernels $F_n$ (which we take to be symmetric under permutations of the arguments).  

For example, using \eqn{fluideqs} and the linear growing mode solutions \eqn{deltasol} and \eqn{linearvel}, we have
\be
S^{(2)}[\kvec ; t] = \int_{\kvec_1 , \kvec_2}^{\kvec} \left\{ \left( 3 f^2 H^2 + H \dot f +f \left( 2 H^2 + \dot H \right)  \right) \alpha_s ( \kvec_1 , \kvec_2 ) - H^2 f^2 \gamma ( \kvec_1 , \kvec_2)  \right\}_t \delta^{(1)}_{\kvec_1} ( t ) \delta^{(1)}_{\kvec_2} ( t ) 
\ee
where here and elsewhere, the subscript $t$ on brackets means that all time arguments inside the brackets not explicitly written are evaluated at $t$, and $\alpha_s$ and $\gamma$ are the symmetrized perturbation theory mixing functions\footnote{Notice that $\gamma ( \kvec_1 , \kvec_2)$ is related to the more commonly used
\be
\beta( \kvec_1 , \kvec_2 ) \equiv   \frac{\hat k_1 \cdot \hat k_2 }{2} \left( \frac{k_1 }{k_2} + \frac{k_2}{k_1} \right)  + \left( \hat k_1 \cdot \hat k_2 \right)^2
\ee
by $\gamma = \alpha_s - \beta$.  }
\be
\alpha_s ( \kvec_1 , \kvec_2 ) \equiv 1 + \frac{\hat k_1 \cdot \hat k_2 }{2} \left( \frac{k_1 }{k_2} + \frac{k_2}{k_1} \right)  \ , \quad \text{and} \quad \gamma( \kvec_1 , \kvec_2 ) \equiv  1- \left( \hat k_1 \cdot \hat k_2 \right)^2 \ . 
\ee
Using the general expression \eqn{generalgf}, we find
\be
F_2 ( \kvec_1 , \kvec_2 ; t ) =  A_\alpha ( t ) \alpha_s ( \kvec_1 , \kvec_2 ) + A_\gamma ( t ) \gamma ( \kvec_1 , \kvec_2 )  \ , 
\ee
where
\begin{align} \label{aalphaintegral}
A_\alpha ( t ) & = \int_0^t d t_1 \, G ( t , t_1 ) \left( 3 f^2 H^2 + H \dot f +f \left( 2 H^2 + \dot H \right)  \right)_{t_1} \frac{D_+(t_1)^2}{D_+(t)^2} \ ,  \\
A_\gamma ( t ) & = - \int_0^t d t_1 \, G ( t , t_1 ) H(t_1)^2 f(t_1)^2 \frac{D_+(t_1)^2}{D_+(t)^2} \ . 
\end{align}

As we will discuss later, the coefficient in front of $\alpha_s$ is  fixed to be $A_\alpha ( t ) = 1$ by the IR symmetries of the fluid system, which means that the integral over the Green's function in \eqn{aalphaintegral} must non-trivially simplify.  The reason that this term is special is because it is the dominant term in $F_2$ when either $k_1 \rightarrow 0$ or $k_2 \rightarrow 0$, i.e. in the IR limit.  We will have much more to say about this in the coming sections.  The coefficient $A_\gamma$, on the other hand, is not fixed by any symmetries, and its value is not restricted in $\Lambda$CDM.  In the EdS cosmology (i.e. with $\om(t) =1$), we have $A_\gamma^{\rm EdS} ( t ) = -2/7$, while for the cosmology given in the Introduction, we have $A_{\gamma} ( t_0 ) = -0.284$, where $t_0$ is the current time.

Generally, we are interested in the statistics of the density field in Fourier space, the most relevant of which, for this work, are the power spectrum $P$, the bispectrum $B$, and the trispectrum $T$.  These are defined by the correlation functions 
\begin{align}
\begin{split} \label{cfndefs}
 \langle \delta_{\kvec_1}(t_1) \delta_{\kvec_2}(t_2) \rangle & = ( 2 \pi)^3 \delta_D ( \kvec_1 + \kvec_2 ) P( k_1  ; t_1,t_2) \ , \\
 \langle \delta_{\kvec_1} (t_1) \delta_{\kvec_2}(t_2) \delta_{\kvec_3}(t_3) \rangle & = ( 2 \pi)^3 \delta_D ( \kvec_1 + \kvec_2  + \kvec_3) \, B ( \kvec_1 , \kvec_2 , \kvec_3 ; t_1,t_2,t_3 )  \ , \\
 \langle \delta_{\kvec_1}(t_1) \delta_{\kvec_2} (t_2)\delta_{\kvec_3} (t_3) \delta_{\kvec_4} (t_4)\rangle  & = ( 2 \pi)^3 \delta_D ( \kvec_1 + \kvec_2  + \kvec_3 + \kvec_4 ) \, T ( \kvec_1 , \kvec_2 , \kvec_3, \kvec_4 ; t_1,t_2,t_3,t_4 )  \ ,
\end{split}
\end{align}
where we have taken the definition of the angle brackets $\langle \cdots \rangle$ to include just the connected pieces.
Translation invariance forces the delta-function structure on the right-hand side of the these expressions.  Rotation invariance makes the power spectrum $P(k_1)$ a function of only the magnitude $k_1$, and in general eliminates three parameters in the dependence of higher point statistics.  For example, it is customary to describe the bispectrum by either the three magnitudes $(k_1 , k_2, k_3)$, or by two magnitudes and a dot product $( k_1 , k_2 , \hat k_1 \cdot \hat k_2 )$.  However, for the purposes of this work, we choose to write the functional dependence of the statistics as in \eqn{cfndefs}.

The expansion \eqn{expanddelta} allows us to compute the various correlation functions in \eqn{cfndefs} perturbatively as well.  First of all, the linear power spectrum $P_{11} ( k)$ is defined by\footnote{To avoid clutter, we will often drop the time arguments from expression like \eqn{cfndefs} when the meaning is clear.  For example, setting the initial conditions $P_{11}(k; t_{\rm in}) = \langle \delta^{\rm in}_{\kvec} \delta^{\rm in}_{\kvec '} \rangle'$, we have, using \eqn{deltasol},  $P_{11}(k; t) = D_+(t)^2 P_{11}(k;  t_{\rm in}) / D_+(t_{\rm in})^2$.  } 
\be \label{p11def}
P_{11} ( k ) \equiv \langle \delta_{\kvec}^{(1)} \delta_{\kvec'}^{(1)} \rangle {}' \ , 
\ee
where the prime on a correlation function means that we strip off the $(2 \pi )^3$ and Dirac delta function which must be present  in \eqn{cfndefs} because of translation invariance.  Next, we can find corrections, called loop corrections, to the linear power spectrum $P_{11}$.  Assuming Gaussian initial conditions, which we assume throughout this work, the expression for the power spectrum up to one loop is
\be
P ( k ) = P_{11} ( k ) + P_{22} ( k ) + P_{13} ( k ) + \dots   \ ,
\ee
where
\be \label{p1loopdefs}
P_{22} ( k ) \equiv \langle \delta^{(2)}_{\kvec} \delta^{(2)}_{\kvec '} \rangle {}' \ , \quad \text{and} \quad P_{13} ( k ) \equiv 2 \langle \delta^{(1)}_{\kvec} \delta^{(3)}_{\kvec'} \rangle {}' \ ,
\ee
and the one-loop contribution is given by $P_{1\text{-loop}} \equiv P_{22} + P_{13}$.\footnote{The explicit expressions for these contributions in the EdS approximation are,
\begin{align}
\begin{split} \label{loopexpressions}
P_{22} ( k )  = 2 \int \momspmeas{q} F_2 ( \qvec , \kvec - \qvec)^2 P_{11} ( q ) P_{11}( | \kvec - \qvec|) \ , \quad \text{and} \quad P_{13}(k)  = 6 P_{11} ( k ) \int \momspmeas{q} F_3 ( \qvec , - \qvec , \kvec) P_{11}(q) \ ,
\end{split}
\end{align}
where the $F_n$ are known kernels \cite{Bernardeau:2001qr}, and, in particular,
\be
 F_2 ( \qvec , \kvec - \qvec ) = \frac{5}{14} + \frac{3 k^2}{28 q^2} + \frac{3 k^2}{28 | \kvec - \qvec |^2}  - \frac{5 q^2}{28 |\kvec - \qvec|^2} - \frac{5 |\kvec - \qvec|^2}{28 q^2} + \frac{k^4}{14 |\kvec - \qvec |^2 q^2}    \ ,
\ee
and
\begin{align}
\begin{split}
F_3 ( \qvec , - \qvec , \kvec ) = & -\frac{97}{1512} + \frac{\diffsq}{24 k^2} + \frac{1195 k^2}{6552 \diffsq} - \frac{19 |\kvec - \qvec|^4 }{504 q^4} + \frac{ \diffsq k^2 }{14 q^4} - \frac{5 k^4}{168 q^4} \\
& - \frac{k^6}{252 \diffsq q^4}  + \frac{211 \diffsq}{1512 q^2} - \frac{| \kvec - \qvec|^4}{72 k^2 q^2} - \frac{187 k^2 }{1512 q^2} -\frac{k^4}{504 \diffsq q^2} \\
&  - \frac{19 q^2}{504 \diffsq} - \frac{q^2}{24 k^2} + \frac{q^4}{72 \diffsq k^2} \ .
\end{split}
\end{align}}
Similar expansions can be done for the bispectrum and trispectrum.  The leading term for the bispectrum $B_t$ ($t$ stands for tree level) is
\be \label{btree}
B_t ( \kvec_1 , \kvec_2  , \kvec_3 ) = \langle \delta_{\kvec_1}^{(2)} \delta_{\kvec_2}^{(1)} \delta_{\kvec_3}^{(1)} \rangle {}'  + \text{2 perms.}
\ee
and the leading term for the trispectrum $T_t$ is given by 
\be
T_t ( \kvec_1 , \kvec_2 , \kvec_3 , \kvec_4 ) = \langle \delta_{\kvec_1}^{(2)} \delta_{\kvec_2}^{(2)} \delta_{\kvec_3}^{(1)}  \delta_{\kvec_4}^{(1)}  \rangle {}'  + \text{5 perms.}  +  \langle \delta_{\kvec_1}^{(3)} \delta_{\kvec_2}^{(1)} \delta_{\kvec_3}^{(1)}  \delta_{\kvec_4}^{(1)}  \rangle {}'  + \text{3 perms.}  \ .
\ee

%
%
%

\section{Consistency relations for dark matter in $\Lambda$CDM} \label{crlcdmreview}

As a setup for our main discussion, we review the arguments leading to the standard consistency relations for dark matter in $\Lambda$CDM.  In this work, we are interested in correlations of the dark-matter overdensity, so we focus on the implications of the consistency relations for these objects below.

%
%
\subsection{Review of non-perturbative statements}

The consistency relations are a set of non-perturbative relationships between soft limits of $(n+1)$-point correlation functions and $n$- or lower-point correlation functions.  They are a result of residual large gauge symmetries\footnote{Large gauge transformations are those that do not vanish at spatial infinity.} in the Newtonian gauge.  The Newtonian-gauge metric \eqn{metric} is invariant under spatial dilations and special conformal transformations {\cite{Creminelli:2012ed, Hinterbichler:2012nm, Hinterbichler:2013dpa, Creminelli:2013mca, Horn:2014rta}}.  These symmetries are relevant for a full relativistic treatment, but in large-scale structure, we deal mostly with the non-relativistic, Newtonian limit, where velocities are much smaller than the speed of light, and typical wavenumbers $k$ are much larger than the Hubble scale.  Then, in the Newtonian limit one is left with two relevant symmetries for the fluid system \eqn{fluideqs}.

The first (which comes from the dilation symmetry) is a shift of the gravitational potential by a time-dependent function,
\be \label{constantshift}
 \Phi ( \xvec, t ) \rightarrow  \Phi ( \xvec , t ) + c_\Phi ( t )  \ , 
\ee
which is unaccompanied by a coordinate change.  This symmetry implies the consistency relation {\cite{Horn:2014rta}}

\be \label{CR1}
 \lim_{q \rightarrow 0} q^2  \frac{ \langle \delta_{\qvec} ( \tau )  \,  \delta_{\kvec_1 } ( t_1) \cdots \delta_{\kvec_n}(t_n) \rangle ' }{P_{11} ( q ; \tau )}  = 0 \ ,
\ee
where here and elsewhere $\tau$ is a time coordinate, but we use a different symbol because it is related to the soft mode.  The second symmetry (which comes from the special conformal symmetry) is a combination of a time-dependent coordinate change and shifts of the fields:
\begin{align}
\begin{split}  \label{transf2}
& \tilde x^i = x^i +  n^i ( t )  \ , \quad {\tilde t  = t } \ , \quad  \tilde v^i ( \tilde x^j , t )  = v^i( x^j , t) + a \dot n^i ( t )  \ , \\
 &\tilde \Phi ( \tilde x^j , t )   = \Phi ( x^j , t)   - a^2( \ddot n^i(t) + 2 H \dot n^i(t) )\tilde x^i  \ , \quad  \tilde \delta ( \tilde x^j , t ) = \delta ( x^j , t ) \ , 
\end{split}
\end{align}
for generic $n^i(t)$.  Equivalently, the transformation \eqn{transf2} acts directly on the equations of motion \eqn{fluideqs} by making the replacements
\begin{align} 
\begin{split} \label{fluidreps}
\partial_i & \rightarrow \partial_i \ , \quad  \partial_t   \rightarrow \partial_t - \dot n^i ( t )   \partial_i  \ ,  \quad  v^i  ( x^j , t ) \rightarrow v^i ( x^j , t) +  a \dot n^i ( t )  \ , \\ 
\Phi ( x^j , t)  & \rightarrow \Phi ( x^j , t)  - a^2 ( \ddot n^i (t ) + 2 H \dot n^i ( t ) ) x^i \ , \quad  \delta (x^j , t) \rightarrow \delta (x^j , t)   \  .
\end{split}
\end{align}
Although the above is a symmetry for any $n^i(t)$, it will generate a physical solution if the time dependence of the transformation matches the time dependence of the $q \rightarrow 0$ limit of a physical solution \cite{Weinberg:2003sw}.  This amounts to $n^i(t)$ solving the linear equation for $D_+(t)$, i.e. \cite{Creminelli:2013mca,Horn:2014rta,Horn:2015dra,Hui:2018cag}
\be
\ddot n^i  + 2 H \dot n^i  - \frac{3 H^2 \om}{2} n^i = 0 \ .
\ee
The symmetry under \eqn{fluidreps}, along with the physical mode condition, implies the consistency relation {\cite{Peloso:2013zw, Kehagias:2013yd, Creminelli:2013mca, Horn:2014rta}}
\be \label{singlecr}
\lim_{q \rightarrow 0}  \frac{ \langle \delta_{\qvec} ( \tau )  \,  \delta_{\kvec_1 } ( t_1) \cdots \delta_{\kvec_n}(t_n) \rangle '  }{P_{11} ( q ; \tau)}  = - \left(\sum_{a=1}^n   \frac{D_+(t_a)}{D_+(\tau)} \frac{ \qvec \cdot \kvec_a }{q^2} \right)  \langle   \delta_{\kvec_1 } (t_1) \cdots \delta_{\kvec_n}(t_n) \rangle '   \ . 
\ee
 Thus, we see that, as $q \rightarrow 0$, terms that grow like $\mathcal{O}(k / q)^2$ (in this work, $\kvec_i$ is always a  short mode, and $\qvec_j$ is always a soft mode), or faster are in general not allowed, and that for equal-time correlators, terms that grow like $\mathcal{O}(k / q)$ or faster are not allowed.  

Similar kinds of arguments can be used to derive consistency relations when multiple external momenta are taken to be soft {\cite{Creminelli:2013poa,Joyce:2014aqa}}.  For example, the double soft-limit consistency relation reads,
\begin{align}
\begin{split} \label{doublesoft}
& \lim_{q_1,q_2 \rightarrow 0}  \frac{ \langle \delta_{\qvec_1} ( \tau_1 ) \delta_{\qvec_2} ( \tau_2)  \,  \delta_{\kvec_1 } ( t_1) \cdots \delta_{\kvec_n}(t_n) \rangle '  }{P_{11} ( q_1 ; \tau_1) P_{11} (q_2 ; \tau_2) }  = \\
& \hspace{1.5in} \left(  \sum_{a=1}^n \frac{D_+(t_a)}{D_+(\tau_1)} \frac{ \qvec_1 \cdot \kvec_a}{q_1^2} \right)   \left( \sum_{b=1}^n \frac{D_+(t_b)}{D_+(\tau_2)} \frac{ \qvec_2  \cdot \kvec_b}{q_2^2} \right) \langle   \delta_{\kvec_1 } (t_1) \cdots \delta_{\kvec_n}(t_n) \rangle '   \ . 
\end{split}
\end{align}

Finally, the same arguments can tell us about loop corrections when the loop momentum is in the deep IR.  We have {\cite{Creminelli:2013poa}} (see also {\cite{Jain:1995kx, Scoccimarro:1995if, Bernardeau:2011vy, Bernardeau:2012aq, Blas:2013bpa, Carrasco:2013sva}}), 
\be \label{irloops}
\langle \delta_{\kvec_1} ( t_1 ) \cdots \delta_{\kvec_n} ( t_n ) \rangle'_{\text{IR loops}}  \approx \exp \left\{ -\frac{1}{2} \int_{\qvec}\,   \left( \sum_{a=1}^n D_+(t_a) \frac{\qvec \cdot \kvec_a}{q^2}  \right)^2 \frac{ P_{11}(q; t_{\rm in})}{D_+(t_{\rm in} )^2}  \right\} \langle \delta_{\kvec_1} ( t_1 ) \cdots \delta_{\kvec_n} ( t_n ) \rangle' \ .
\ee
 By expanding the exponential to $N$-th order, we find the effect of $N$ soft loops on the correlation function.

We would like to stress that these relationships are non-perturbative, and in fact have an even broader applicability than for only the dark-matter density.  {We present a derivation of these results using the background wave argument in \appref{crreviewapp}.}  Strictly speaking, these results are only correct when $q$ is taken much smaller than all energy scales in the linear power spectrum, of which the most relevant in our universe is the BAO scale {\cite{Senatore:2014via,Baldauf:2015xfa}}.  In the correlation function,\footnote{The correlation function is the Fourier transform of the power spectrum, and is explicitly given by 
\be
\xi ( r ) = \int_{\kvec}  P ( k) e^{i \kvec \cdot \rvec} = \int_0^\infty d k  \frac{k^2}{2 \pi^2} \frac{\sin kr}{kr} P(k) \ .
\ee}
the BAO show up as a peak near $\ellbao \approx 110 \unitsr$ of width $\sigma \approx 10 \unitsr$ (this means that in the power spectrum, there is an oscillatory feature with frequency $\Delta k_{\rm osc} \sim 2 \pi \ellbao^{-1}$ and support $\Delta k_{\rm width} \sim 1/ \sigma$, see \figref{wnwfig}).  Thus, the consistency relations are valid for $q \ll 2 \pi \ellbao^{-1}$.  We discuss this further in \secref{bispsubsec}.

%
\subsection{Perturbative argument for the IR with exact time dependence} \label{pertirsec}

In this section, we show how to derive the above relations in perturbation theory, since this method will ultimately be used later when we study DHOST theories.  Additionally, we work with the full Green's function to compute the time dependence of the fields, as opposed to the commonly employed EdS approximation, since this approach will be relevant for our later discussion.  As we will see, our approach will not depend on the solution $\delta^{(1)}$ being a growing mode.  

We would like to solve \eqn{fluideqs} for $\delta^{(n)}$ at leading order for $q/k \rightarrow 0$.  To proceed, we could simply iteratively solve \eqn{fluideqs} as normal, and then take the IR limit.  However, we can use our knowledge of the symmetries of the system to simplify the job (again, this point of view will be more helpful later).  

First, we notice that invariance of the equations of motion under \eqn{fluidreps} implies a relationship between terms with time derivatives and certain non-linear terms involving the velocity.  For example, a time derivative on $\delta$ must occur in the combination, commonly known as the convective derivative,
\be
\dot \delta + a^{-1} v^i \partial_i \delta  \ ,
\ee
to be invariant.  The same goes for two derivatives on $\delta$, which must occur in the combination
\be
\ddot \delta + a^{-1} ( \dot v^i \partial_i \delta + 2 v^i \partial_i \dot \delta - H v^i \partial_i \delta ) +a^{-2}v^i v^j \partial_i \partial_j \delta \ . 
\ee
These are the leading IR terms that appear in the equation of motion.  This can be seen because they have no spatial derivatives on the velocity (which is necessary because of the symmetry \eqn{fluidreps}), and the long-wavelength velocity \eqn{linearvel} is the only object in the equations of motion that can introduce inverse powers of $q$.  This means that we can directly write the leading IR contributions to the equation of motion as
\be \label{ireom}
\ddot \delta + 2 H \dot \delta - a^{-2} \partial^2 \Phi \approx - a^{-1} ( \dot v^i \partial_i \delta + 2 v^i \partial_i \dot \delta + H v^i \partial_i \delta) - a^{-2} v^i v^j \partial_i \partial_j \delta \ ,
\ee
where here and in the rest of this work, the symbol $\approx$ means that the relationship is valid to leading order as $q/k \rightarrow 0$.  We have shown, given the left-hand side of \eqn{ireom}, that the terms on the right-hand side must be present in order for the equation to be Galilean invariant.  But are these the only terms allowed?  We could imagine adding other Galilean invariant non-linear terms, such as $\delta ( \dot \delta + a^{-1} v^i \partial_i \delta)$, to the right-hand side.  However, one can quickly see that these will be sub-dominant in the IR limit because they do not have enough powers of $v^i$.  Thus, the form of \eqn{ireom} in the IR limit is in this sense unique.

This equation can then be solved directly using the Green's function \eqn{gfdef}, which gives the formal solution
\begin{align}
\begin{split} \label{np2eq}
& \delta^{(n+2)} ( t )  \approx - \int^t d \ttt \, \bar G ( t , \ttt ) \,  a( \ttt )^{-1} \Big\{  \dot v^i_{(1)} \partial_i \delta^{(n+1)} + 2 v^i_{(1)} \partial_i \dot \delta^{(n+1)} + H v^i_{(1)} \partial_i \delta^{(n+1)}  \\
& \hspace{4in} + a^{-1} v^i_{(1)} v^j_{(1)} \partial_i \partial_j \delta^{(n)}   \Big\}_{\ttt}  \ , 
\end{split}
\end{align}
where we have suppressed the space coordinate $\xvec$.  After some manipulations, and using properties of the Green's function, the solution to this equation in Fourier space is given by (see \appref{gfapp} for details), 
 \be  \label{fssol}
 \delta^{(n+1)}_{\kvec} \approx \int_{\qvec_1, \cdots, \qvec_n} \frac{1}{n!} \left( \prod_{a=1}^n  \frac{\qvec_a \cdot \kvec}{q_a^2}  \delta^{(1)}_{\qvec_a} \right) \delta^{(1)}_{\kvec - \sum_b \qvec_b} \ .
 \ee
For example, we have
\be
\delta^{(2)}_{\kvec} \approx \int_{\qvec_1} \frac{\qvec_1 \cdot \kvec}{q^2_1} \delta^{(1)}_{\qvec_1} \delta^{(1)}_{\kvec  - \qvec_1}  \ , \quad \text{and} \quad \delta^{(3)}_{\kvec} \approx \int_{\qvec_1 , \qvec_2} \half \frac{\qvec_1 \cdot  \kvec}{q_1^2} \frac{ \qvec_2 \cdot  \kvec}{q_2^2} \delta^{(1)}_{\qvec_1} \delta^{(1)}_{\qvec_2} \delta^{(1)}_{\kvec - \qvec_1 - \qvec_2} \ .
\ee
In terms of the perturbation theory kernels $F_n$, this means
\be
F_2 ( \qvec_1 , \kvec) \approx \frac{1}{2} \frac{\qvec_1 \cdot \kvec}{q_1^2} \ , \quad \text{and} \quad F_3 ( \qvec_1 , \qvec_2 , \kvec) \approx \frac{1}{6} \frac{\qvec_1 \cdot \kvec}{q_1^2} \frac{\qvec_2 \cdot \kvec}{q_2^2} \ ,
\ee
for $q_{1,2} \ll k$. {Notice, crucially, that the Green's function from \eqn{np2eq} has dropped out of the solution \eqn{fssol}.  The way that this happens is quite non-trivial, so we present the proof in \appref{gfapp}.

The form of the solution \eqn{fssol} is not an accident, and in fact could have been derived in a much more straightforward way, avoiding the Green's function altogether.  The invariance of the equations of motion under \eqn{transf2} for a time-dependent $n^i(t)$ tells us that if we give $n^i(t)$ a weak spatial dependence, the solution for $\delta$, at the lowest order in derivatives, is given by a coordinate transformation.  Promoting $ n^i ( t ) \rightarrow n^i_L ( \xvec , t)$ (here and elsewhere the subscript $L$ is to remind us that the function is a long-wavelength field), we choose 
\be
a \dot n^i_L ( \xvec , t ) = - v^i_{(1)} ( \xvec , t) \ .
\ee
Then, by construction, the equation of motion for $\tilde \delta$ in the tilde coordinates does not contain the leading IR non-linear terms, i.e. the terms on the right-hand side of \eqn{ireom}, and so $\tilde \delta$ satisfies the linear equations to leading order in the IR.  Taking the field in the tilde coordinates to be the linear solution $\tilde \delta ( \tilde x^i , t) = \delta^{(1)} ( \tilde x^i , t)$, we have, to leading order in derivatives,
\be \label{coordtransfexp}
\delta ( \xvec , t ) \approx  \delta^{(1)} ( \xvec + \vec n_L ( \xvec , t), t ) \ . 
\ee
In Fourier space, this becomes (dropping the time dependence),
\be \label{coordtransf1}
\delta_{\kvec}  \approx \int d^3 x \, e^{-i \kvec \cdot \xvec} \int_{\kvec'} e^{i \kvec' \cdot ( \xvec + \vec{n}_L ( \xvec ) )} \delta^{(1)}_{\kvec'} \ .
\ee
Then, using the linear continuity equation, we have 
\be \label{chooseni}
 n^i_L ( \xvec  ) =  \frac{\partial_i \delta^{(1)} ( \xvec )}{\partial^2} =  - i \int_{\qvec} \frac{q^i}{q^2} \delta^{(1)}_{\qvec} e^{i \qvec \cdot \xvec} \ . 
\ee
and plugging this into \eqn{coordtransf1} gives  
\be \label{coordtransf2}
\delta_{\kvec}  \approx \int d^3 x \int_{\kvec'} e^{i ( \kvec'- \kvec )\cdot \xvec}  \exp \left\{ \int_{\qvec} \frac{\qvec \cdot \kvec'}{q^2} \delta^{(1)}_{\qvec} e^{i \qvec \cdot \xvec}  \right\} \delta^{(1)}_{\kvec'} \ .
\ee
Expanding the exponential that contains $\delta^{(1)}_{\qvec}$ then gives the solution \eqn{fssol}, which we refer to as the standard $\Lambda$CDM solution.  {This derivation makes it clear that the Green's function in \eqn{np2eq} has to cancel, which we show explicitly in \appref{gfapp}.} This form of the leading IR solution is the basis of the consistency relations.

%
%
%

\section{DHOST in the IR} \label{dhostsec}

\subsection{Gravitational equations and linear solutions}

To find the equations in the gravitational sector for DHOST, we expand the action \eqn{EFTaction} in terms of the metric and scalar field perturbations $\pi$ and keep only terms with the highest number of derivatives per field, which are those relevant in the Newtonian limit.  After doing that and varying the action with respect to the fields, we obtain \cite{Dima:2017pwp, Crisostomi:2019yfo, Langlois:2017mxy},
\begin{align} \label{phieq1}
\begin{split}
& \frac{a^2 \bar \rho \, \delta }{2 M^2} = C_1\partial^2 \pi - \frac{c_8}{4}  \partial^2 \dot \pi + \frac{c_6}{2} \partial^2 \Phi   + \frac{c_4}{4}  \partial^2 \Psi   +  \frac{1}{4} \left[  \frac{b_2}{a^2} Q_2 [\pi,\pi ] + \frac{c_8}{a^2} \partial_i \left( \partial_j \pi \partial_i \partial_j \pi \right) \right] \ , 
\end{split}
\end{align}
\begin{align} \label{psieq1}
\begin{split}
&  0 = C_2 \partial^2 \pi - \frac{c_7}{4}  \partial^2 \dot \pi + \frac{c_4}{4}  \partial^2 \Phi  + \frac{c_5}{2}  \partial^2 \Psi  +  \frac{1}{4} \left[  \frac{b_3}{a^2} Q_2 [\pi,\pi ] + \frac{c_7}{a^2} \partial_i \left( \partial_j \pi \partial_i \partial_j \pi \right) \right] \ , 
\end{split}
\end{align}
and 
\begin{align}\label{pieq1}
\begin{split}
   0  = & C_3  \partial^2 \pi +  C_4 \partial^2 \dot \pi + \frac{c_9}{2} \partial^2 \ddot \pi  +  \frac{c_1}{4}  \partial^2 \Phi     + \frac{c_8}{4}  \partial^2 \dot \Phi   +  \frac{c_2}{4}  \partial^2 \Psi + \frac{c_7}{4}  \partial^2 \dot \Psi   \\
&  +  \frac{1}{4 a^2}  Q_2 [ \pi , b_1\pi+ 2b_2 \Phi + 2b_3 \Psi ]   - \frac{1}{4 a^2} \partial_i \left[ \partial_i \pi  \, \partial^2 ( c_7 \Psi + c_8 \Phi  + 2 c_9 \dot \pi) \right] \\
&   + \frac{ H c_9 - C_4 }{2 a^2}   \partial^2 \left( \partial \pi \right)^2  - \frac{c_9}{2 a^2 } \partial^2 (  \partial_i \pi \partial_i \dot \pi  )  - \frac{b_2 + b_3}{4 a^4} Q_3 [ \pi , \pi , \pi] + \frac{c_9}{4 a^4} \partial_i \left( \partial_i \pi \partial^2 (\partial \pi )^2 \right)  \ ,
\end{split}
\end{align}
where we have defined 
\begin{align}
\begin{split}
Q_2 [ \varphi_a , \varphi_b]  & \equiv \varepsilon^{ikm} \varepsilon^{jlm} \partial_i \partial_j \varphi_a \partial_k \partial_l \varphi_b  \ , \\
 Q_3[ \varphi_a, \varphi_b  , \varphi_c ]  &  \equiv  \varepsilon^{ikm} \varepsilon^{jln} \partial_i \partial_j \varphi_a \partial_k \partial_l \varphi_b \partial_m \partial_n \varphi_c \ ,
\end{split}
\end{align}
with 
\be
\varphi_a \equiv \{ \Phi, \Psi, \pi\}  \;,
\ee
and $C_1, \dots, C_4$, 
 $c_1, \ldots, c_9$, and $b_1, b_2, b_3$ are time-dependent coefficients that depend on the parameters of the action, reported in \appref{coefficientsapp}.  Although we focus on the leading IR terms in the coming sections, we have presented the full equations of motion above for completeness.\footnote{{While we focus on the IR structure of the above equations, we note that \cite{Hirano:2020dom} found that the UV contribution from loops does not satisfy momentum conservation.}}

We seek a perturbative solution to \eqns{phieq1}{pieq1} in powers of $\delta$. Thus, we will expand the fields $\varphi_a$ as
\be
\varphi_a = \varphi_a^{(1)} + \varphi_a^{(2)} + \varphi_a^{(3)} + \dots \;,
\ee 
 where each perturbative piece is proportional to the relevant number of powers of  $\delta^{(1)}$, i.e. $\varphi_a^{(n)} \sim [\delta^{(1)}]^n$.  In this work, we will solve up to third order, and we start with the linear solutions.

As discussed in \cite{Crisostomi:2019yfo} (see also \cite{Hirano:2018uar,Crisostomi:2017pjs,Hirano:2019nkz,Hirano:2019scf}), the linear solutions have the following form
\begin{align}
\begin{split} \label{linearsols}
 a^{-2} \partial^2 \varphi_a^{(1)}  = \mu_{\varphi_a} \delta^{(1)} + \nu_{\varphi_a} \dot \delta^{(1)} + \sigma_{\varphi_a} \ddot \delta^{(1)}  \  .
\end{split}
\end{align}
We have supplied the expressions for the time dependent $\mu_{\varphi_a}$, $\nu_{\varphi_a}$, and $\sigma_{\varphi_a}$ functions in terms of the parameters in the 
field equations
in \appref{linearsolssec}, but we note, however, that 
\be
\sigma_\pi = 0 \ .
\ee   
Horndeski theories have $\sigma_{\varphi_a} = \nu_{\varphi_a} = 0$, and $\Lambda$CDM has $\mu_\Phi = \mu_\Psi = \bar \rho / ( 2 \mpl^2) = 3 \om H^2 / 2 $.

Combining \eqn{linearsols} with the fluid equations \eqn{fluideqs}, we have the linear evolution equation for $\delta^{(1)}$, 
\be \label{lineareom}
\ddot \delta^{(1)} + \bar \nu_\Phi \dot \delta^{(1)} - \bar \mu_\Phi \delta^{(1)} = 0 \ , 
\ee
where for future convenience, we have defined 
\be
\bar \nu_\Phi \equiv \frac{2 H - \nu_\Phi}{1 - \sigma_\Phi} \ , \quad \bar \mu_\Phi \equiv \frac{\mu_\Phi }{1 - \sigma_\Phi}  \ . 
\ee
The linear equation \eqn{lineareom} has two solutions, one growing, $D_+(t)$, and one decaying, $D_- (  t )$.  We focus on the growing mode solution so we write the solution for $\delta^{(1)}$ as \eqn{deltasol}.  Looking at \eqn{fluideqs}, this means that the linear solution for the velocity can be written as in \eqn{linearvel}.

{Once we have the linear solution $D_+$, this allows us to write the linear solutions \eqn{linearsols} as 
\begin{align}
\begin{split} \label{Ldefs1}
a^{-2} \partial^2 \varphi_a^{(1)} =  L_{\varphi_a} \delta^{(1)} \ , 
\end{split}
\end{align}
where
\begin{align}
\begin{split} \label{Ldefs}
L_{\varphi_a}  = \mu_{\varphi_a} + H f \nu_{\varphi_a} + (H^2 f^2 + H \dot f + \dot H f ) \sigma_{\varphi_a}  \ . 
\end{split}
\end{align}

%

\subsection{Symmetries and leading IR terms}

To start, we note that the constant shift symmetry \eqn{constantshift} and associated consistency relation \eqn{CR1} remain in the DHOST system.  The symmetry under $\Phi ( \xvec , t ) \rightarrow \Phi ( \xvec , t ) + c_\Phi ( t )$ means that the correlation of short modes $\delta_{\kvec_n}$ in the background of a long wavelength $\Phi_L$ does not depend on the value of $\Phi_L$, i.e.
\be \label{constantshiftcoorfn}
\langle \delta_{\kvec_1} \cdots \delta_{\kvec_n} \rangle_{\Phi_L} = \langle \delta_{\kvec_1} \cdots \delta_{\kvec_n} \rangle_{\Phi_L + c_\Phi}  \ .
\ee
In particular, we can expand the dependence of the correlation function on derivatives of $\Phi_L$ as (for $q \ll k_i$),
\be 
\langle \delta_{\kvec_1} \cdots \delta_{\kvec_n} \rangle_{\Phi_L} = \langle \delta_{\kvec_1} \cdots \delta_{\kvec_n} \rangle_{0} + \int_{\qvec} \frac{\delta \langle \delta_{\kvec_1} \cdots \delta_{\kvec_n} \rangle_{\Phi_L} }{\delta ( q^i \Phi_{\qvec} )} \Bigg|_0 \, q^i \Phi_{\qvec} + \dots
\ee
where the term proportional to the derivative with respect to $\Phi_{\qvec}$ is absent because of \eqn{constantshiftcoorfn}.  {Assuming that there is a {well defined derivative expansion of $\langle \delta_{\kvec_1} \cdots \delta_{\kvec_n} \rangle_{\Phi_L}$ with respect to $\Phi_L$}, the functional derivative shown above is finite.} Then, taking the correlation with another long wavelength $\Phi_{\qvec}$, we find
\be \label{expandcorrfn}
\langle \Phi_{\qvec}\, \delta_{\kvec_1} \cdots  \delta_{\kvec_n} \rangle \approx  - q^i P_{\Phi} ( q ) \frac{\delta \langle \delta_{\kvec_1} \cdots \delta_{\kvec_n} \rangle_{\Phi_L} }{\delta ( q^i \Phi_{\qvec } )} \Bigg|_0  + \mathcal{O} \left( q^2 P_\Phi ( q ) \right)  \ , 
\ee
where $P_{\Phi} ( \qvec ) \equiv \langle \Phi_{\qvec} \, \Phi_{\qvec_1} \rangle'$.\footnote{Note that the functional derivative $ \delta \langle \delta_{\kvec_1} \cdots \delta_{\kvec_n} \rangle_{\Phi_L} / \delta ( q^i \Phi_{\qvec } ) |_0 $ is proportional to $\delta_D ( \sum_a \kvec_a - \qvec)$, so that the delta functions work out in \eqn{expandcorrfn}. }  Then, we can use the linear equation for $\partial^2 \Phi$ in \eqn{Ldefs1} to write 
\be
\lim_{q \rightarrow 0 } q^2 \frac{\langle \delta_{\qvec} \, \delta_{\kvec_1} \cdots \delta_{\kvec_n} \rangle }{P_{11} ( q)} = q^i L_{\Phi} a^2  \frac{\delta \langle \delta_{\kvec_1} \cdots \delta_{\kvec_n} \rangle_{\Phi_L} }{\delta ( q^i \Phi_{\qvec } )} \Bigg|_0   \rightarrow 0 \ ,
\ee
which confirms that \eqn{CR1} is valid in DHOST theories.

Next, we consider the Galilean transformations.  The equations in the gravitational sector \eqns{phieq1}{pieq1} are invariant under the following coordinate change and shifts of the fields:  
\begin{align}
\begin{split} \label{transf1}
  \tilde x^i &=  x^i +  \xi^i ( t )  \ ,  \quad {\tilde t =  t  }  \ , \quad  \tilde \delta ( \tilde x^j ,\tilde  t )  = \delta ( x^j , t )  \ , \quad \tilde \varphi_a ( \tilde x^j , \tilde t )  = \varphi_a ( x^j , t) + b^i_{\varphi_a} ( t ) \tilde  x^i  \ . \\
\end{split}
\end{align}
Equivalently,  they are invariant under the  replacements
\begin{align}
\begin{split}  \label{newtransf1}
\partial_i  \rightarrow \partial_i  \;, \qquad \partial_t   \rightarrow \partial_t  - \dot \xi^i ( t )   \partial_i  \ , \quad \varphi_a  (x^j , t) \rightarrow \varphi_a (x^j , t)  + b^i_{\varphi_a} ( t ) x^i \ , \quad \delta ( x^j,t)  \rightarrow \delta ( x^j , t)  \;.
\end{split}
\end{align} 

In Horndeski theories, which have $\alphaH = \beta_1 = 0$, this transformation is a symmetry for arbitrary functions $\xi^i ( t )$, $b^i_\Phi ( t )$,  $b^i_\Psi ( t )${,} and  $b^i_\pi ( t )$, as can be easily verified, since we have $c_6=c_7=c_8=c_9=C_4=0$ and all fields have at least two spatial derivatives in \eqns{phieq1}{pieq1}.

In DHOST theories, on the other hand, when $\alphaH \neq 0$ or $\beta_1 \neq 0$, the equations \eqns{phieq1}{pieq1} are only invariant under \eqn{transf1} if 
\be \label{bipi}
b^i_\pi ( t ) = - a^2 \dot \xi^i ( t ) \ ,
\ee 
i.e. if the transformation of $\pi$ is fixed by the coordinate transformation.  This is because the equations contain terms with time derivatives and less than two spatial derivatives.  Under \eqn{newtransf1}, a time derivative of a field generates a term involving that field and $\dot \xi^i$, which can only be canceled by a higher order term involving $\partial_i \pi$ if \eqn{bipi} holds.

As discussed in \cite{Crisostomi:2019vhj}, when combined with the fluid equations \eqn{fluideqs}, the system has an overall Galilean invariance, i.e. when setting $\xi^i ( t ) = n^i(t)$, and $b^i_{\Phi} ( t ) = - a^2 ( \ddot n^i ( t ) + 2 H \dot n^i (t ) )$.  Defining the scalar field velocity
\be
\label{vipi}
v^i_{\pi } \equiv -a^{-1} \partial_i \pi  \;,
\ee
we see that the transformations of $v^i$ and $v^i_\pi$ are the same under this transformation.  This means that the relative velocity, defined by
\be
\Delta v^i \equiv v^i - v^i_\pi \ , 
\ee
is invariant under the Galilean transformation, which in turn means that we cannot use the Galilean transformation to eliminate the effect of the long-wavelength relative velocity, i.e. we cannot construct a physical adiabatic mode.  Thus, the consistency relations are violated by terms proportional to the relative velocity.  Incidentally, this means that DHOST theories that have $\Delta v^i = 0$ do not violate the consistency relations (see \eqn{deltavzero} for the explicit condition).  In Horndeski theories, on the other hand, $b^i_\pi ( t )$ is not fixed to be related to the transformation of $v^i$, and so one can construct the physical adiabatic mode, and therefore the consistency relations are satisfied \cite{Crisostomi:2019vhj} (see also \cite{Cusin:2017wjg} for a proof that the one-loop power spectrum is IR safe).  

Although the consistency relations are generically violated in DHOST theories, we can use the symmetries \eqn{fluidreps} and \eqn{newtransf1} to compute this violation in terms of the parameters in the linear equations of motion, and explicitly show how they are proportional to the relative velocity $\Delta v^i$. 

In particular, the symmetry \eqn{transf1} means that in the solution to \eqns{phieq1}{pieq1}, i.e. for the scalar potentials $\Phi$, $\Psi$, and $\pi$ in terms of the overdensity $\delta$, time derivatives of fields must be accompanied by non-linear terms containing the velocity $v^i_\pi$.  For example, a time derivative on $\delta$ must occur in the combination
\be
\dot \delta + a^{-1} v^i_\pi \partial_i \delta  \ 
\ee
to be invariant.  The same goes for two derivatives on $\delta$, which must occur in the combination
\be
\ddot \delta + a^{-1} ( \dot v^i_\pi \partial_i \delta + 2 v^i_\pi \partial_i \dot \delta - H v^i_\pi \partial_i \delta ) +a^{-2}v^i_\pi v^j_\pi \partial_i \partial_j \delta \ . 
\ee
Then, writing the non-linear solutions for the gravitational fields as 
\be \label{varphiexpand}
a^{-2} \partial^2 \varphi_a  =  \mu_{\varphi_a} \delta + \nu_{\varphi_a} \dot \delta + \sigma_{\varphi_a} \ddot \delta + a^{-2} \partial^2 \varphi_a^{\rm NL} \ , 
\ee
we see immediately that, in order for our solution be invariant under \eqn{transf1}, we must have 
\be \label{phinlsol}
a^{-2} \partial^2 \varphi_a^{\rm NL} \approx a^{-1} \left(  \nu_{\varphi_a}  v^i_{\pi} \partial_i \delta   + \sigma_{\varphi_a} \left(   \dot v^i_{\pi} \partial_i \delta + 2 v^i_{\pi} \partial_i \dot \delta - H v^i_{\pi} \partial_i \delta  +a^{-1}v^i_{\pi} v^j_{\pi} \partial_i \partial_j \delta \right)   \right)  \ ,
\ee
We stress here that we did not have to solve the full non-linear equations: along with the linear equations, the symmetry \eqn{transf1} determines the leading non-linear expressions in the IR.  In the same way as discussed below \eqn{ireom} for the fluid equations, the form of the IR solution \eqn{phinlsol} is unique in the sense that, given the linear part of the solution in \eqn{varphiexpand}, \eqn{phinlsol} gives the unique non-linear extension which is invariant under \eqn{transf1}.  

%
%
\subsection{Perturbative solutions in the IR}

As discussed in \cite{Crisostomi:2019vhj}, we can use the symmetries \eqn{transf2} and \eqn{transf1} to easily compute the leading IR contributions to the perturbative solutions $\delta^{(2)}$ and $\delta^{(3)}$.  Plugging the IR expression for $\partial^2 \Phi$ from \eqn{varphiexpand} into the IR expression for the fluid from \eqn{ireom}, we have, after some rearranging, 
\begin{align}
\begin{split} \label{IReom}
\ddot \delta + \bar \nu_\Phi \dot \delta - \bar \mu_\Phi \delta \approx & - a^{-1} \left( 2 v^i \partial_i \dot \delta + \dot v^i \partial_i \delta - H v^i \partial_i \delta \right) - a^{-2} v^i v^j \partial_i \partial_j \delta  - \bar \nu_\Phi a^{-1} v^i \partial_i \delta \\
& - \frac{ a^{-1} }{1-\sigma_\Phi} \Big[ \nu_\Phi   \Delta v^i \partial_i \delta +  \sigma_\Phi \left( 2 \Delta v^i \partial_i \dot \delta +  \Delta \dot v^i \partial_i \delta - H \Delta v^i \partial_i \delta \right) \\
& \hspace{.8in} + \sigma_\Phi a^{-1} \left( v^i v^j - v^i_\pi v^j_\pi \right) \partial_i \partial_j \delta   \Big]  \ . 
\end{split}
\end{align}
As discussed above, the IR non-linear extensions of the fluid equations \eqn{ireom} and the gravitational fields \eqn{phinlsol} are each unique, determined by the respective symmetries \eqn{transf2} and \eqn{transf1}.  Thus, the final expression \eqn{IReom} is unique given the way that the gravitational potential $\partial^2 \Phi$ enters \eqn{ireom}; if the coupling of the gravitational potentials to matter changes, \eqn{IReom} could change as well.  That said, given our assumption of minimal coupling, \eqn{IReom} is the unique IR non-linear extension of the linear equations that preserves the symmetries \eqn{transf2} and \eqn{transf1}.

As we show in \appref{gfapp}, if we set $\Delta v^i = 0$, the solution to the above equation is \eqn{assume}, which is the same as the $\Lambda$CDM solution \eqn{coordtransf2}.  This makes sense, since when $\Delta v^i = 0$, the leading IR solution is given simply by the coordinate transformation \eqn{coordtransfexp}.  As we will show in \secref{observablessec}, the presence of the terms proportional to $\Delta v^i$ in \eqn{IReom} violates the consistency relations.

First, we find $\delta^{(2)}$, which solves
\begin{align} \label{d2ireom}
\ddot \delta^{(2)} + \bar \nu_\Phi \dot \delta^{(2)} - \bar \mu_\Phi \delta^{(2)} \approx & - a^{-1} \Big[ 2 v^i_{(1)} \partial_i \dot \delta^{(1)} + \dot v^i_{(1)} \partial_i \delta^{(1)} - (H  - \bar \nu_\Phi ) v^i_{(1)} \partial_i \delta^{(1)} \Big]  \\
& - \frac{ a^{-1} }{1-\sigma_\Phi} \Big[\left(  \nu_\Phi  - H \sigma_\Phi \right) \Delta v^i_{(1)} \partial_i \delta^{(1)} +  \sigma_\Phi \left( 2 \Delta v^i_{(1)} \partial_i \dot \delta^{(1)} +  \Delta \dot v^i_{(1)} \partial_i \delta^{(1)} \right)  \Big]  \ . \nonumber
\end{align}
Applying the Green's function, we see from \appref{gfapp} that the first line above gives the standard $\Lambda$CDM contribution.  The second line cannot be further simplified, so it gives a generic contribution which is proportional to the linear relative velocity.  Using \eqn{vipi} and \eqn{Ldefs1}, the linear velocity of the scalar field and the linear relative velocity can be written as
\be
v^i_{\pi (1)} = - a L_\pi \frac{\partial_i \delta^{(1)}}{\partial^2} \ , \quad \text{and} \quad \Delta v^i_{(1)} = - a L_{\Delta v} \frac{\partial_i \delta^{(1)}}{\partial^2} \ , \quad \text{for} \quad L_{\Delta v } \equiv Hf - L_\pi \ .
\ee
The final expression is \cite{Crisostomi:2019vhj}
\begin{align} \label{delta2solde}
\delta^{(2)} ( t ) \approx   A_\alpha ( t )  \frac{\partial_i \delta^{(1)} ( t ) }{\partial^2} \partial_i \delta^{(1)} ( t )  \ ,
\end{align}
where 
\begin{align} \label{Aalphadef}
A_\alpha ( t )  = 1 +  \int^t_0 d \ttt \, \bar G(t , \ttt ) K_2 ( \ttt) \frac{D_+(\ttt)^2}{D_+(t)^2} \ , \quad K_2 \equiv   \frac{    \nu_\Phi L_{\Delta v}    +  \sigma_\Phi ( 3 H  f L_{\Delta v} + \dot L_{\Delta v}    )  }{1 - \sigma_\Phi  }   \ . 
\end{align}
 We show an example plot of $A_\alpha ( t ) -1$ in \figref{Aminus1fig}.

 \begin{figure*}[t] 
\centering 
 \includegraphics[width=0.65\textwidth]{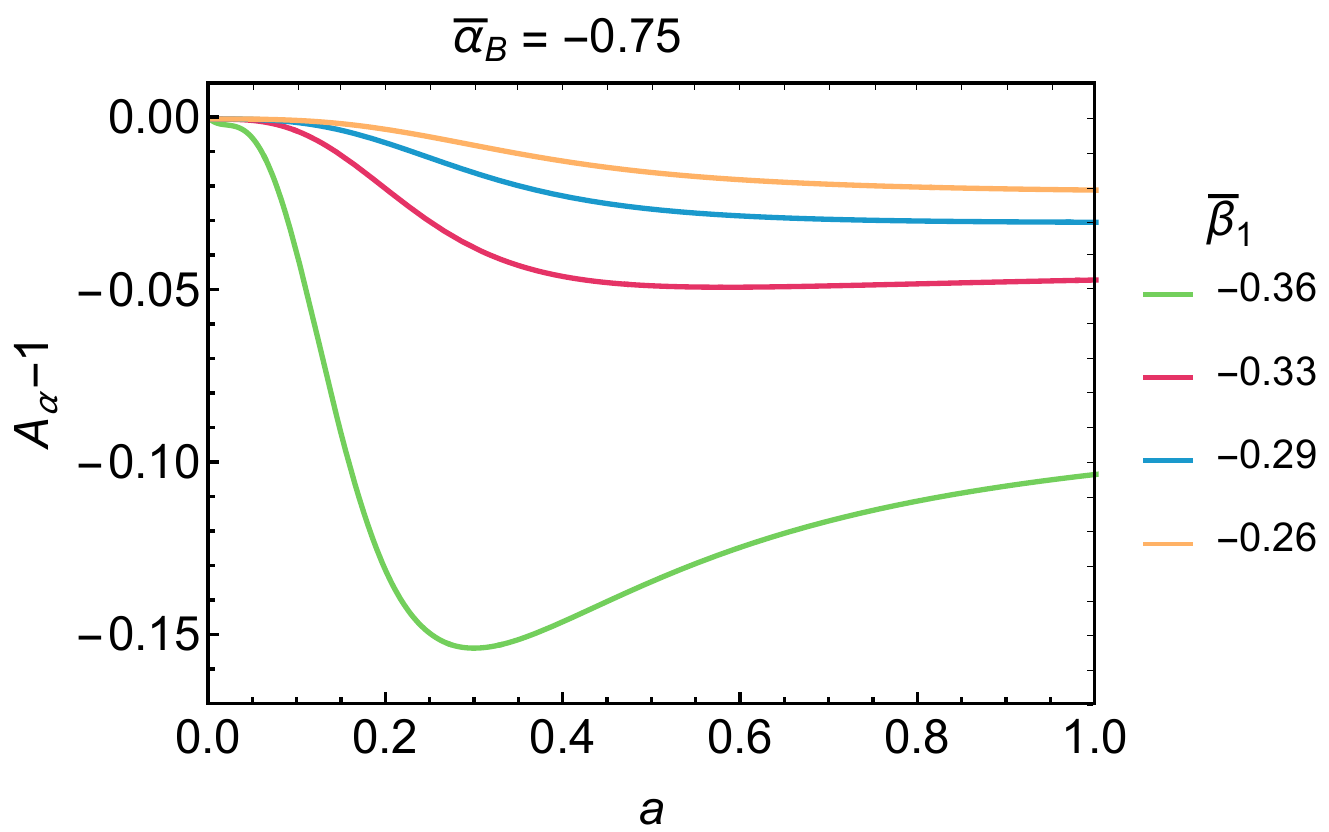} 
\caption{ Function $A_\alpha - 1$, see \eqn{Aalphadef}, as a function of the scale factor $a$ for various values of $\beta_1$ and a fixed $\alphaB$.  The background evolution has been chosen to be the one of $\Lambda$CDM, i.e.~the Hubble rate is given by $H ( a )  = H_0 \sqrt{a^{-3} \Omega_{\rm m , 0} + 1 - \Omega_{\rm m , 0}} $, the matter evolution is given by {$\Omega_{\rm m} ( a )   = \Omega_{\rm m , 0} / ( \Omega_{\rm m , 0} + a^3 ( 1 - \Omega_{\rm m , 0} )) $}, and we have taken $\Omega_{\rm m,0} = 0.281$ as the current value of the fractional matter density. (In the numerical calculation, the Hubble rate always appears in the combination $H/H_0$ so that the curves are independent of the value of $H_0$.)  
We  parametrize the time dependence of the  EFT of dark energy parameters as $\alphaB ( a ) = \bar \alpha_{\rm B} ( 1 - \Omega_{\rm m} ( a ) )$ and $\beta_1 ( a ) = \bar \beta_{1} ( 1 - \Omega_{\rm m} ( a ) )$, where $ \bar \alpha_{\rm B}$ and $\bar \beta_1$ are constants.
The other EFT parameters, for simplicity, are chosen such that the model leaves the gravitational wave speed, amplitude, and decay unaffected (see e.g.~\cite{Crisostomi:2019yfo} for a discussion), i.e.~$\alphaT=\alphaM = 0$ and $\alphaH = - 2 \beta_1$.
Moreover, we  only plot values of $\bar \alpha_{\rm B}$ and $\bar \beta_{1}$ for which $ \alpha c_s^2 > 0$, as required by the absence of ghost and gradient instability (see e.g.~\cite{Gleyzes:2014rba}).    } \label{Aminus1fig}
\end{figure*}

\begin{figure*}[t] 
\centering 
\includegraphics[width=0.65\textwidth]{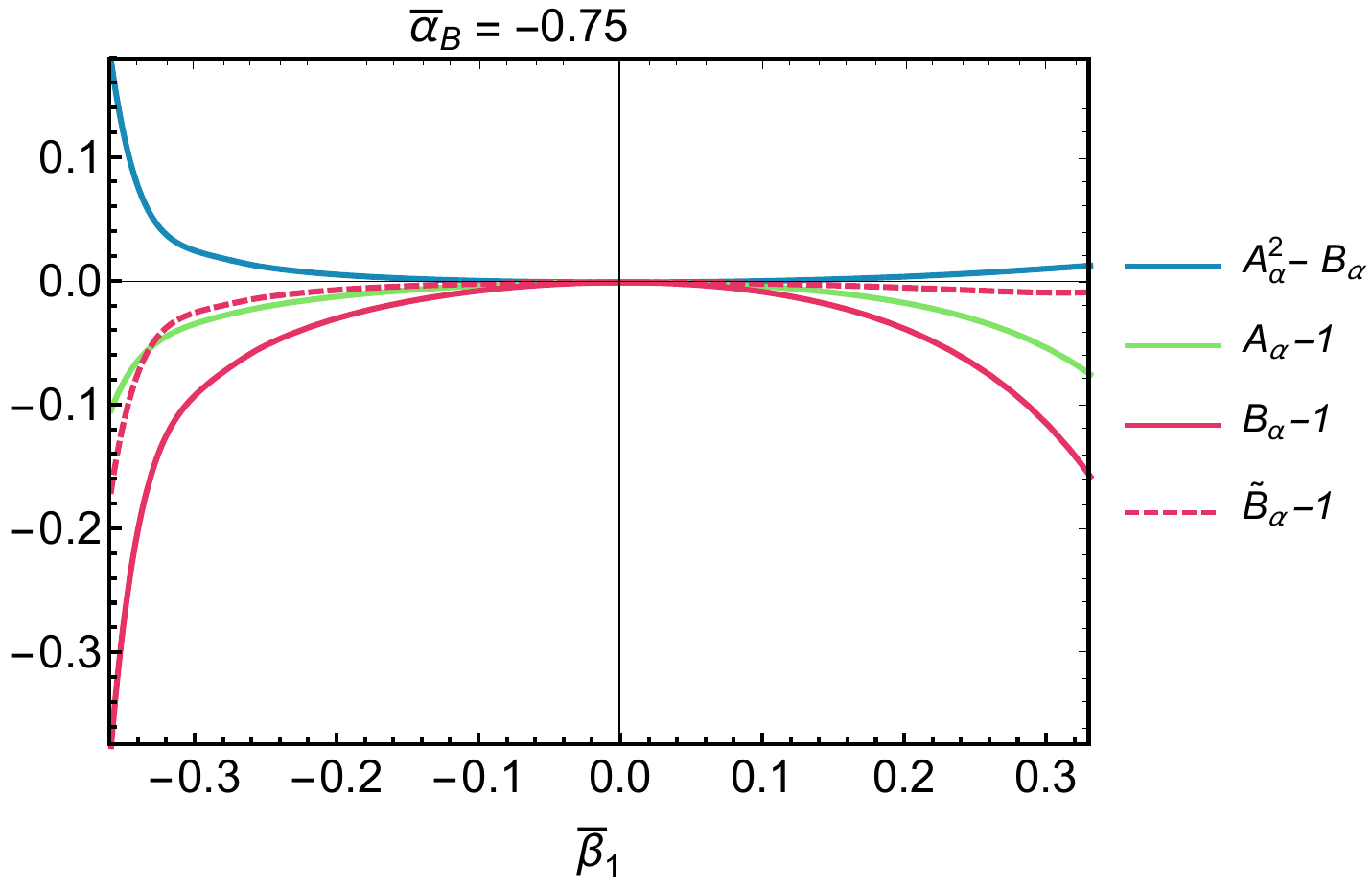}
\caption{ Various functions related to the violation of the consistency relations in DHOST theories.  We plot the modification of the $\delta^{(2)}$ solution ($A_\alpha -1$, see \eqn{Aalphadef}), the modification of the $\delta^{(3)}$ solution ($B_\alpha -1$, see \eqn{Balphadef}), the modification of the one-loop power spectrum and tree-level trispectrum ($A_\alpha^2 - B_\alpha$, see \eqn{p1looplimit}  and \eqn{trispectrumlimit} ), and a related function $\tilde B_\alpha - 1$ (see \eqn{tildebalphafn}), as a function of $\bar \beta_1$ for fixed $\bar \alpha_{\rm B} = - 0.75$, at $ a=1$.  The parameterizations of cosmological and EFT parameters are given in the caption of \figref{Aminus1fig}.  
{Notice that $A_\alpha = 1 $ and $B_\alpha = 1$ in Horndeski theories, i.e.~for $\beta_1 = 0$, as expected. }    } \label{Asfigure}
\end{figure*}

Next, using the second-order solution above, we can solve for the third-order field.  The equation for $\delta^{(3)}$ in the IR is 
\begin{align} \label{delta3eqn}
\ddot \delta^{(3)} + \bar \nu_\Phi \dot \delta^{(3)} - \bar \mu_\Phi \delta^{(3)} \approx & - a^{-1} \Big[ \dot v^i_{(1)} \partial_i \delta^{(2)} + 2 v^i_{(1)} \partial_i \dot \delta^{(2)} +( H - \bar \nu_\Phi)  v^i_{(1)} \partial_i \delta^{(2)} + a^{-1} v^i_{(1)} v^j_{(1)} \partial_i \partial_j \delta^{(1)}   \Big] \nonumber  \\
& - \frac{ a^{-1} }{1-\sigma_\Phi} \Big[ (  \nu_\Phi - H \sigma_\Phi )   \Delta v^i_{(1)} \partial_i \delta^{(2)} +  \sigma_\Phi \left( 2 \Delta v^i_{(1)} \partial_i \dot \delta^{(2)} +  \Delta \dot v^i_{(1)} \partial_i \delta^{(2)}  \right)   \nonumber  \\
& \hspace{.8in} + \sigma_\Phi a^{-1} \left( v^i_{(1)} v^j_{(1)} - v^i_{\pi (1)} v^j_{\pi (1)} \right) \partial_i \partial_j \delta^{(1)}   \Big]  \ .
\end{align}
Now, we apply the Green's function.  When we plug the $\Lambda$CDM solution for $\delta^{(2)}$ into the first line of \eqn{delta3eqn}, we obtain the $\Lambda$CDM solution for $\delta^{(3)}$ (see \appref{gfapp}).  For the rest of the terms, we simply integrate over the Green's function to obtain
\begin{align} \label{delta3sol}
\delta^{(3)} ( t ) \approx   B_\alpha ( t )    \half   \frac{ \partial_i  \delta^{(1)} ( t)}{\partial^2}   \frac{ \partial_j  \delta^{(1)} ( t)}{\partial^2}  \partial_i \partial_j   \delta^{(1)} ( t)   \ , 
\end{align}
where 
\begin{align} \label{Balphadef}
B_\alpha ( t ) =  1 + \int_0^t d \ttt \, \bar G ( t , \ttt)  K_3 (\ttt) \frac{D_+(\ttt)^3}{D_+(t)^3}
\end{align}
and we have defined
\begin{align}
\begin{split} \label{k3def}
K_3 & =  \frac{2}{1-\sigma_\Phi} \Big[  (\ldv^2 + \ldvd) \sigma_\Phi + \ldv (\nu_\Phi +3 H f \sigma_\Phi )  + 2 \Delta \dot A_\alpha  \left( f H ( 1 - \sigma_\Phi) + \ldv \sigma_\Phi  \right)  \\
&  \quad \quad \quad \quad \quad   + \Delta A_\alpha \left(  \mu_\Phi + 4 f^2 H^2 (1 - \sigma_\Phi)  + \ldv (\nu_\Phi + 5 f H \sigma_\Phi ) + \ldvd \sigma_\Phi    \right)    \Big] \ ,
\end{split}
\end{align}
and $\Delta A_\alpha \equiv A_\alpha - 1$.  This is the most straightforward way to obtain the solution, but as we show later in \secref{powerspectrumsec}, the result can be further simplified by first changing coordinates in the equation of motion.  We show various functions related to the violation of the consistency relations in \figref{Asfigure}.

We would like to stress here how, after knowing the manipulation of the Green's function which gives the standard $\Lambda$CDM result, we are able to easily determine the solutions in the IR without going through the full equations of motion.  As a final note, we point out that a subset of DHOST theories, those with $L_{\Delta v}= 0$, i.e.  
\be \label{deltavzero}
H f - \mu_\pi - H f \nu_\pi = 0 \ ,
\ee
do not violate the consistency relations.  However, it appears quite difficult to satisfy this condition, since it would mean that the DHOST parameters are functionally dependent on the background evolution $\om ( t ) $ and the linear growth rate $f$.

%
%
%

\section{Observables} \label{observablessec}

%
\subsection{Bispectrum} \label{bispsubsec}

We first look at the squeezed limit of the equal-time matter bispectrum.  Using \eqn{delta2solde}, we have
\be \label{bispecsl}
\lim_{q \rightarrow 0} \frac{B_t ( \qvec, \kvec , \kvec_1) }{P_{11} ( q ) } \approx - A_\alpha \frac{\qvec \cdot ( \kvec + \kvec_1 )}{q^2} P_{11}(k)  \ ,
\ee
which starts at $\mathcal{O}((k/q)^0)$ because $\kvec + \kvec_1 = - \qvec$.  
Therefore, there is no $k/q$ enhancement of the full bispectrum in the squeezed limit $q \, \ellbao \ll 1$.  However, the vanishing of the right-hand side of \eqn{bispecsl} is not a consequence of the consistency relations  but simply of the symmetry of the bispectrum under exchange of the two arguments $\vec k$ and $\vec k_1$ {(and translation invariance, i.e. that $\kvec + \kvec_1 + \qvec = 0$).}  As discussed in \cite{Crisostomi:2019vhj}, though, there will generically be an enhancement in the squeezed limit of the bispectrum of different species (for example, different types of galaxies).

\begin{figure*}[t] 
\centering 
\hspace{-.3in} \includegraphics[width=0.5\textwidth]{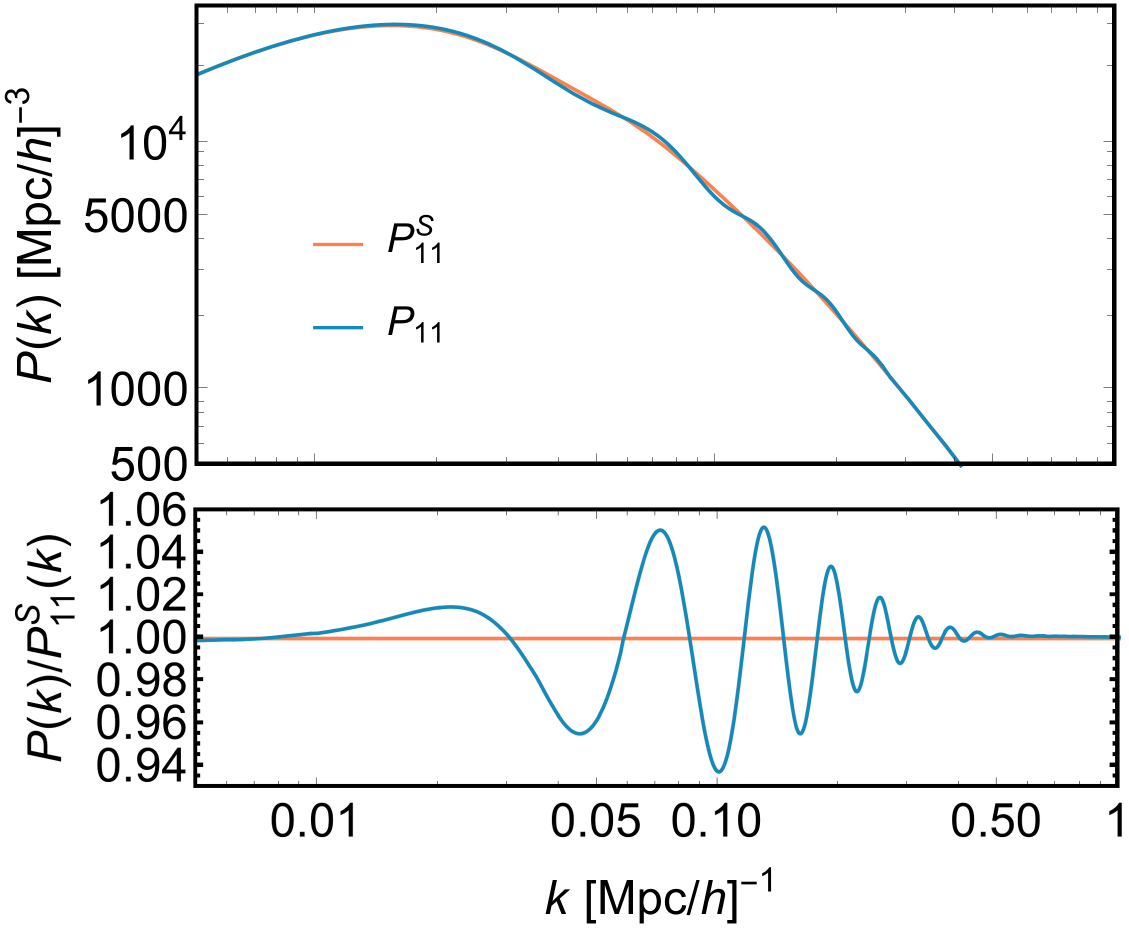}  \includegraphics[width=0.5\textwidth]{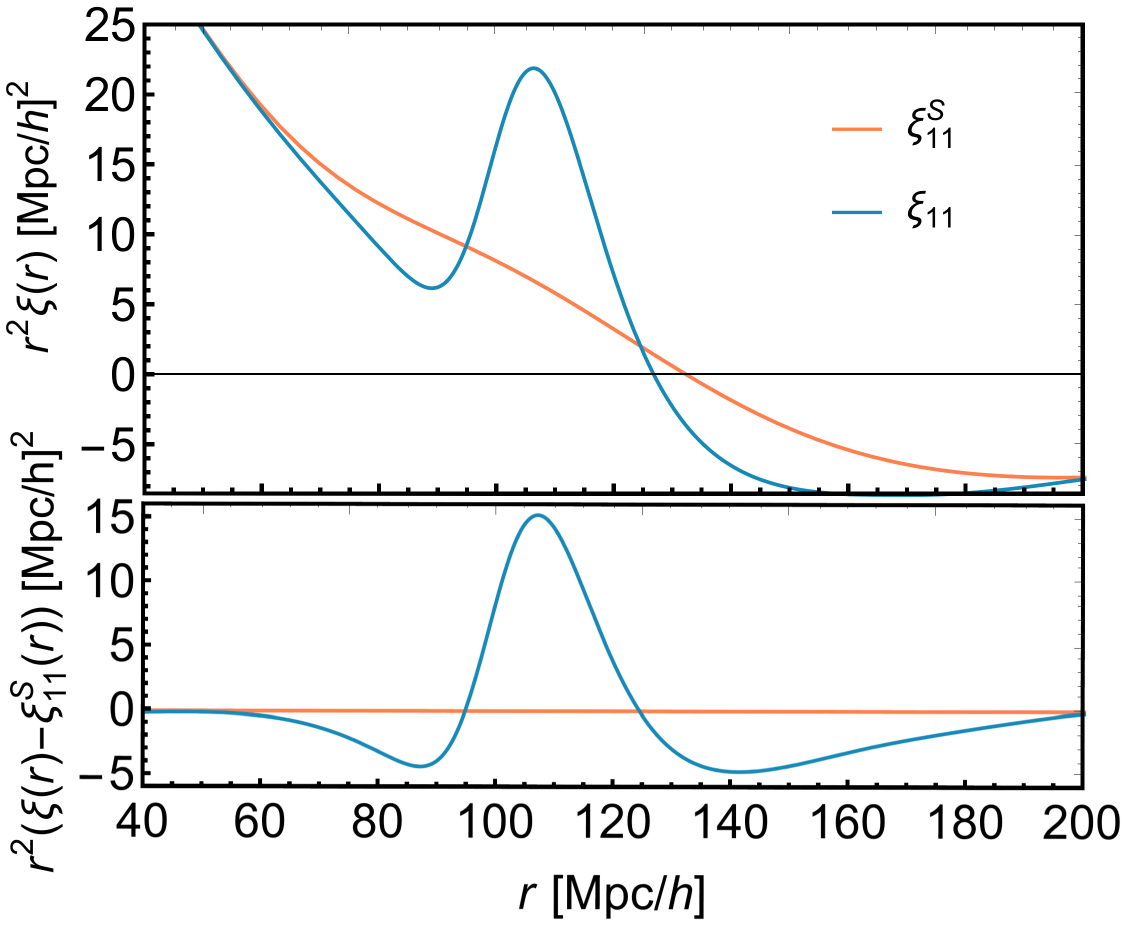}
\caption{  The BAO feature in the power spectrum (left panel) and correlation function (right panel) for linear theory in $\Lambda$CDM.  In blue is the relevant linear quantity ($P_{11}$ and $\xi_{11}$), while in orange is the smoothed version of that quantity ($P_{11}^{\rm s}$ and $\xi^{\rm s}_{11}$) (see for example \cite{Vlah:2015zda} for one way to define the smoothed quantities).  In the bottom panels, we show the residual effects.  } \label{wnwfig}
\end{figure*}

In order to find an effect in the bispectrum of all the same species, we have to go to the next order in $q/k$.  
Using \eqn{delta2solde} and expanding to the next order, we find,
\be \label{bispecslsub}
\lim_{q \rightarrow 0} \frac{B_t ( \qvec, \kvec  - \qvec /2, -\kvec - \qvec / 2) }{P_{11} ( q ) } \approx 0 - A_\alpha \frac{\qvec \cdot \kvec}{q^2} \frac{\qvec \cdot \kvec}{k} \frac{\partial P_{11} ( k )}{\partial k} + \mathcal{O} \left(  P_{11}(k) \right) \ .
\ee
The zero above comes from \eqn{bispecsl}, which is valid at $\mathcal{O}(k/q)$.  The next term shown is naively of order $(k/q)^0$, and so would be of the same order as other terms that have been neglected.  However, if the term shown in \eqn{bispecslsub} is large compared to the neglected terms for other reasons, it can dominate, giving again a contribution which is computable using only IR properties (i.e. depends only on $A_\alpha$). This is precisely what happens in the presence of the BAO, where the term shown in \eqn{bispecslsub} can be the dominant contribution to the oscillatory part of the bispectrum (see for example \cite{Baldauf:2015xfa} for a discussion in $\Lambda$CDM).  In \figref{wnwfig}, we show the BAO feature for linear theory.

To see how the term in \eqn{bispecslsub} can be the dominant contribution to oscillatory features, it is useful to consider computing the smooth part of the power spectrum separately from the oscillatory part (see, for example, \cite{Baldauf:2015xfa,Vlah:2015zda}).  To do that, we split the linear power spectrum into a smooth part $P^{\rm s}_{11}$ and an oscillatory part $P_{11}^{\rm osc}$, as in \figref{wnwfig},\footnote{In this work, we ignore any ambiguity in splitting the linear power spectrum (see for example \cite{Lewandowski:2018ywf}). }
\be
P_{11} ( k ) = P^{\rm s}_{11} ( k ) +  \epsilonosc P^{\rm osc}_{11} ( k )  \ ,
\ee
where $\epsilonosc$ is a small parameter describing the amplitude of the oscillations, which from \figref{wnwfig} we see is $\epsilonosc \sim 0.06$.  For the smooth part, which is approximately scale free, we have
\be
\frac{\partial P_{11}^{\rm s} ( k ) }{\partial k } \sim \frac{P_{11}^{\rm s} ( k)}{k}  \ , 
\ee
while for the oscillatory part, we have 
\be \label{poscder}
\frac{\partial P_{11}^{\rm osc} ( k )}{\partial k} \sim \frac{ \ellbao}{2\pi} P_{11}^{\rm osc} ( k )  \  .
\ee
Then, we can compute the bispectrum in powers of $\epsilonosc$, writing 
\be
B_t = B_t^{\rm s} + \epsilonosc B_t^{\rm osc} + \mathcal{O}(\epsilonosc^2) , 
\ee
and matching the first power of $\epsilonosc$ in \eqn{bispecslsub}, we have
\be \label{bispecslsubosc}
\lim_{q \rightarrow 0} \frac{B_t^{\rm osc} ( \qvec, \kvec  - \qvec / 2, - \kvec - \qvec / 2) }{P_{11}^{\rm s} ( q ) } \approx - A_\alpha \frac{\qvec \cdot \kvec}{q^2} \frac{\qvec \cdot \kvec}{k} \frac{\partial P_{11}^{\rm osc} ( k )}{\partial k} + \mathcal{O} \left(  P^{\rm osc}_{11}(k) \right) \ .
\ee
Now, because of \eqn{poscder}, the term that we have included in \eqn{bispecslsubosc} will dominate over the terms that we have neglected if 
\be
k \, \ellbao \gg 1 \ ,
\ee
which is satisfied for typical wavenumbers of interest.  In $\Lambda$CDM, we have $A_\alpha = 1$, and so this contribution to the oscillatory part of the bispectrum is universal, fixed by the equivalence principle.  However, we see that DHOST theories change this contribution.

\begin{figure*}[t] 
\centering 
\hspace{-.3in} \includegraphics[width=.8\textwidth]{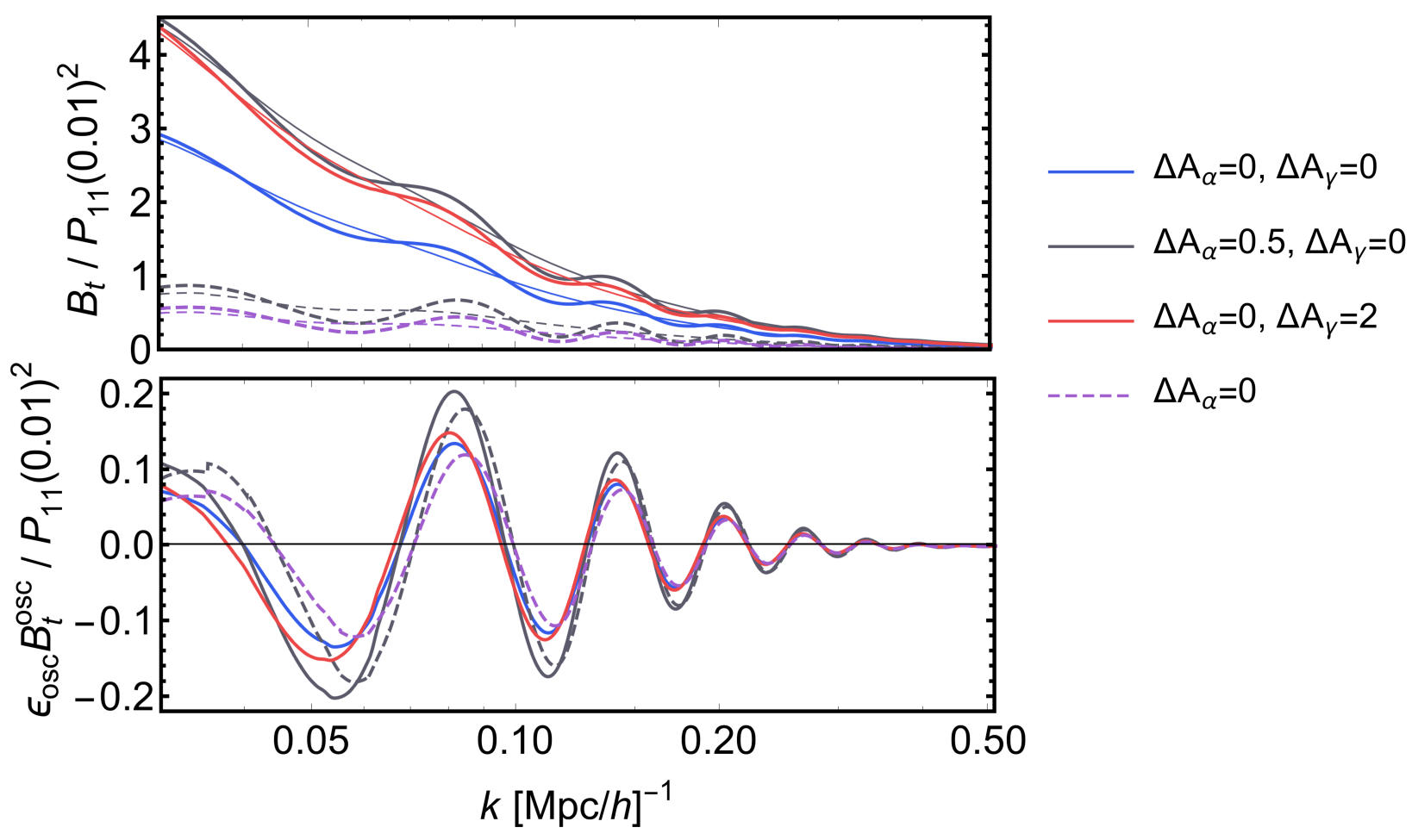} 
\caption{ We plot various bispectra in the configuration $(\qvec , \kvec - \qvec / 2 , - \kvec - \qvec / 2)$ for $q = 0.01 \unitsk$ and $\hat q \cdot \hat k = 0.9$, for various values of $\Delta A_\alpha$ and $\Delta A_\gamma$.  Solid lines are the full bispectrum \eqn{fullbisp}, while dashed lines are the dominant oscillatory contributions \eqn{bispecslsubosc}.  We have also plotted the associated smooth bispectra with thin lines.  In the bottom panel, we plot the residual $\epsilonosc B_t^{\rm osc} = B_t - B_t^{\rm s}$.  In the bottom panel, we see that there are two different limiting behaviors, the purple and grey dashed lines, corresponding to $A_\alpha = 0$ and $A_\alpha = 0.5$ respectively.  The red and blue solid curves have different values of $A_\gamma$, but they have the same limiting behavior (the purple dashed line) because they have the same value of $A_\alpha$.  In all cases, the full bispectra approach the correct limiting behavior \eqn{bispecslsubosc} for $k / q \gg 1$.  We note that the large value of $\Delta A_\gamma$ needed to produce a visible difference between the red and blue curves in the plot is due to the fact that the contribution is proportional to $A_{\gamma} ( 1 - (\hat q \cdot \hat k )^2)$, and we have chosen $\hat q \cdot \hat k = 0.9$ so that \eqn{bispecslsubosc} would be the dominant contribution.    } \label{bslplot}
\end{figure*}

In \figref{bslplot}, we show how this term enters the bispectrum.  We write the full bispectrum as in \cite{Crisostomi:2019vhj},
\be \label{fullbisp}
B_t ( \qvec , \kvec_1 , \kvec_2 ) = 2 F_2 ( \qvec , \kvec_1 ) P_{11} ( q ) P_{11} (  k_1 )  + \text{2 perms.} \ ,
\ee 
where
\be
F_2 ( \qvec , \kvec) = A_\alpha \alpha_s ( \qvec , \kvec) + A_\gamma \gamma ( \qvec , \kvec ) \ . 
\ee
To describe the deviations from $\Lambda$CDM, we write
\be
A_\alpha = 1 + \Delta A_\alpha \ , \quad \text{and}  \quad A_\gamma = A_\gamma^{\Lambda \text{CDM}} + \Delta A_\gamma \ , 
\ee
where for the cosmology used in this paper, we have $A_\gamma^{\Lambda \text{CDM}} = - 0.284$.   

In \figref{bslplot}, we plot the full bispectrum \eqn{fullbisp} and the dominant oscillatory part \eqn{bispecslsubosc} for various choices of $\Delta A_\alpha$ and $\Delta A_\gamma$.  We see that the oscillatory parts of the full bispectra all approach the dominant oscillatory contribution \eqn{bispecslsubosc} for $k / q \gg 1$, and that this limit depends on $A_\alpha$, but not $A_\gamma$.  Thus, the measurement of a different size of the oscillations in the squeezed limit would be a clear signal for the violation of the consistency relations, which could have its origin in DHOST theories with $A_\alpha \neq 1$.  Of course, to actually measure this effect, one needs to compute the galaxy or halo bispectrum, which we turn to next.

%
%
\subsection{Bias expansion} \label{biassec}

In this section, we are interested in the bias expansion relevant for computing the galaxy bispectrum, so we consider terms up to second order.  Following \cite{McDonald:2009dh}, we write the overdensity of galaxies (or any other tracer) $\delta^g$ as a derivative expansion of the underlying dark-matter field.  Since $\delta^g$ is a scalar under the overall Galilean transformation (i.e. \eqn{transf2} and \eqn{transf1} with $\xi^i = n^i$ and $b^i_\Phi = - a^2 ( \ddot n^i + 2 H \dot n^i)$), we write the bias expansion in terms of Galilean scalars.  As discussed, $\Delta v^i$ is a scalar, and so it should be included in the bias expansion.  This means that the new terms that can be added are proportional to, at first order, $\partial_i \Delta v^i$, and at second order, $\Delta v^i \partial_i \delta$, $\delta \partial_i \Delta v^i$, and $\Delta v^i \Delta v^i$.  However, the terms containing $\partial_i \Delta v^i$ are already allowed in the $\Lambda$CDM bias expansion, so they do not provide new terms here.  Thus, we have, up to second order,
\be
\delta^g = b_{g,1} \delta + b_{g,2} \delta^2 + b_{g,3} \frac{\partial_i \partial_j \delta }{\partial^2}\frac{\partial_i \partial_j \delta }{\partial^2} + b_{g,4} \Delta v^i \Delta v^i + \frac{b_{g,5}}{a H} \Delta v^i \partial_i \delta  + \dots
\ee
where all fields above are evaluated at the same coordinates $(\xvec , t)$, and the $b_{g,i}$ are time-dependent dimensionless coefficients.\footnote{For the sake of brevity, we are ignoring the stochastic bias and subtleties related to non-locality in time \cite{Senatore:2014eva}.  Similar terms regarding the relative velocity have appeared in the context of baryons (see for example \cite{Schmidt:2016coo}).   }  Since the new terms must vanish in Horndeski, we must have that $b_{g,4} = 0$ and $b_{g,5}=0$ when $\nu_\Phi = \sigma_\Phi = 0$.

Relative to the size of the other contributions, we expect the $(\Delta v )^2$ term to be a relativistic correction, since in momentum space it scales as an extra factor of $H^2 / k^2$ with respect to, for example, the $\delta^2$ term.  In any case, we are interested in the squeezed limit, and the new term which is relevant in that regime is the one proportional to $b_{g,5}$, which is absent in $\Lambda$CDM.  As mentioned above, if we correlate different tracers, there can be an enhanced effect in the squeezed limit of the bispectrum, proportional to the difference in the bias coefficients.  For this discussion, though, we are interested in the signal in the bispectrum of a single tracer.  Following the discussion in the previous section, there will be no enhancement in the broadband of the bispectrum, so we will look at the BAO.  Defining
\be
B^g ( \kvec_1 , \kvec_2, \kvec_3) \equiv \langle \delta^g_{\kvec_1} \delta^g_{\kvec_2} \delta^g_{\kvec_3} \rangle ' \ , 
\ee
we have
\be \label{bispecslsuboscgal}
\lim_{q \rightarrow 0} \frac{B_t^{g, \text{osc}} ( \qvec, \kvec  - \qvec / 2, - \kvec - \qvec / 2) }{P_{11}^{g,\text{s}} ( q ) } \approx -A_\alpha^g  \frac{\qvec \cdot \kvec}{q^2} \frac{\qvec \cdot \kvec}{k} \frac{\partial P_{11}^{g,\text{osc}} ( k )}{\partial k} + \mathcal{O} \left(  P_{11}^{g,\text{osc}}(k) \right) \ ,
\ee
where we have defined 
\be
A_\alpha^g \equiv   \frac{ A_\alpha}{b_{g,1}}  - \frac{b_{g,5} L_{\Delta v}}{b_{g,1}^2 H} \ ,
\ee
and $P_{11}^g ( k ) = b_{g,1}^2 P_{11} ( k ) $, and the smooth and oscillatory pieces are defined as in \secref{bispsubsec}.

In theories that satisfy the consistency relations, we have $A_\alpha^g = 1 / b_{g,1}$, and so measuring the size of the oscillations in the squeezed limit, as in \figref{bslplot}, is a direct measurement of the linear bias $b_{g,1}$ (see \cite{Marinucci:2019wdb} for a confirmation in simulations).  Unfortunately, this means that the contribution to \eqn{bispecslsuboscgal} from DHOST appears degenerate with the linear bias, and that only a more thorough analysis of multiple observables, so that one can independently determine $b_{g,1}$, will be able to disentangle the DHOST contribution from the bias contribution.  We leave this kind of study for future work.

%
\subsection{Power spectrum and trispectrum} \label{powerspectrumsec}

The violation of the consistency relations also has consequences for the IR contribution to the one-loop power spectrum.
 Using \eqn{delta2solde} and \eqn{delta3sol} in \eqn{p1loopdefs}, we have,
\be \label{p1looplimit}
P_{1\text{-loop}} ( k ) \approx P_{1\text{-loop}}^{\rm IR} (k ) \equiv  ( A_\alpha^2 - B_\alpha ) P_{11} ( k ) \int_{q \lesssim k}  \frac{d^3 q}{(2 \pi)^3} \left( \frac{\qvec \cdot \kvec}{q^2} \right)^2 P_{11} ( q )  \ .
\ee 
  This contribution is absent in $\Lambda$CDM due to the equivalence principle  \cite{Jain:1995kx, Scoccimarro:1995if, Peloso:2013zw, Carrasco:2013sva}, but is present here in DHOST because of the non-zero large-scale relative velocity.\footnote{ {We note that, for late-time dark-energy models (which is what we are mostly concerned with in this paper), the linear power spectrum goes as $P_{11}(q ) \sim q^{n_s}$ in the IR, where $n_s \approx 0.971$ is the tilt of the primordial power spectrum.  This ensures that the integral in \eqn{p1looplimit} is convergent, and that perturbation theory can be well-defined.  }}

To see how this new term affects perturbation theory, we first introduce the expansion parameters of the Eulerian loop expansion.  These are given by {\cite{Senatore:2014via,Baldauf:2015xfa,Tassev:2013rta}},
\be \label{epsilons}
\esl ( k ) \equiv  k^2 \int_0^k \frac{d^3 q}{(2 \pi)^3} \frac{P_{11} ( q ) }{q^2}  \ , \, \, \edl ( k ) \equiv \int_0^k \frac{d^3 q }{(2 \pi)^3} P_{11} ( q )  \ , \, \text{and} \, \, \esg ( k ) \equiv k^2 \int_k^\infty  \frac{d^3 q}{(2 \pi)^3} \frac{P_{11} ( q ) }{q^2} \ .
\ee
As we can see from \eqn{epsilons}, $\esl$ is due to IR displacements, $\esg$ is due to UV displacements, and $\edl$ is due to IR density fluctuations.  In this notation, we have
\be \label{p1loopirdef}
P_{1\text{-loop}}^{ \rm IR} (k)  = ( A_\alpha^2 - B_\alpha) P_{11}(k) \,  \frac{\esl ( k )}{3} \ .
\ee
As shown in \figref{epsilonsfig}, $\esl$ is significantly larger than $\edl$.  In standard $\Lambda$CDM, the terms proportional to $\esl$ cancel between $P_{22}$ and $P_{13}$ because of the equivalence principle, and so the Eulerian expansion is in powers of $\edl$, which is indeed much smaller on the scales of interest.\footnote{{The large IR displacements do, however, have an effect on the computation of the BAO, {which we describe in more detail in \secref{irresumsec}}.  In order to correctly describe the BAO in perturbation theory one must resum the effects of the long-wavelength displacements (see for example \cite{Senatore:2014via, Matsubara:2007wj, Lewandowski:2015ziq, Senatore:2017pbn, Lewandowski:2018ywf, Baldauf:2015xfa, Vlah:2015sea, Vlah:2015zda, Blas:2016sfa, delaBella:2017qjy, Ivanov:2018gjr}).  Being an IR effect, this does not change the broadband convergence properties of the perturbative expansion.}}  Thus, the presence of \eqn{p1loopirdef} can affect the validity of the perturbative expansion, which can now be controlled by the IR displacements.  

In order to describe the new terms in the loop expansion, we introduce two expansion parameters.  The first are the DHOST parameters $\nu_\Phi / H $ and $\sigma_\Phi$, which are both proportional to $\alphaH$ or $\beta_1$.  When these parameters are non-zero, we have terms in the gravitational field equations \eqns{phieq1}{pieq1} that have one and three spatial derivatives, which is the hallmark of DHOST theories.  In what follows, in order to count the order of perturbations, we use $\lambda_D$ to stand for $\nu_\Phi / H $ and $\sigma_\Phi$.  The other expansion parameter, which we denote as $\lambda_{\Delta v}$, is the relative velocity, i.e. $ \lambda_{\Delta v} \sim L_{\Delta v} / H$.  The violation of the consistency relations is proportional to both of these parameters, so that both have to be non-zero in order to have an effect.  This makes sense, since even in Horndeski theories we generically have $\lambda_{\Delta v} \neq 0$, but there is no violation of the consistency relations because $\lambda_D = 0$.  On the other hand, we can have DHOST theories, $\lambda_D \neq 0$, that do not violate the consistency relations if $\lambda_{\Delta v } = 0$, see \eqn{deltavzero}.

\begin{figure*}[t] 
\centering 
\hspace{-.3in} \includegraphics[width=.57\textwidth]{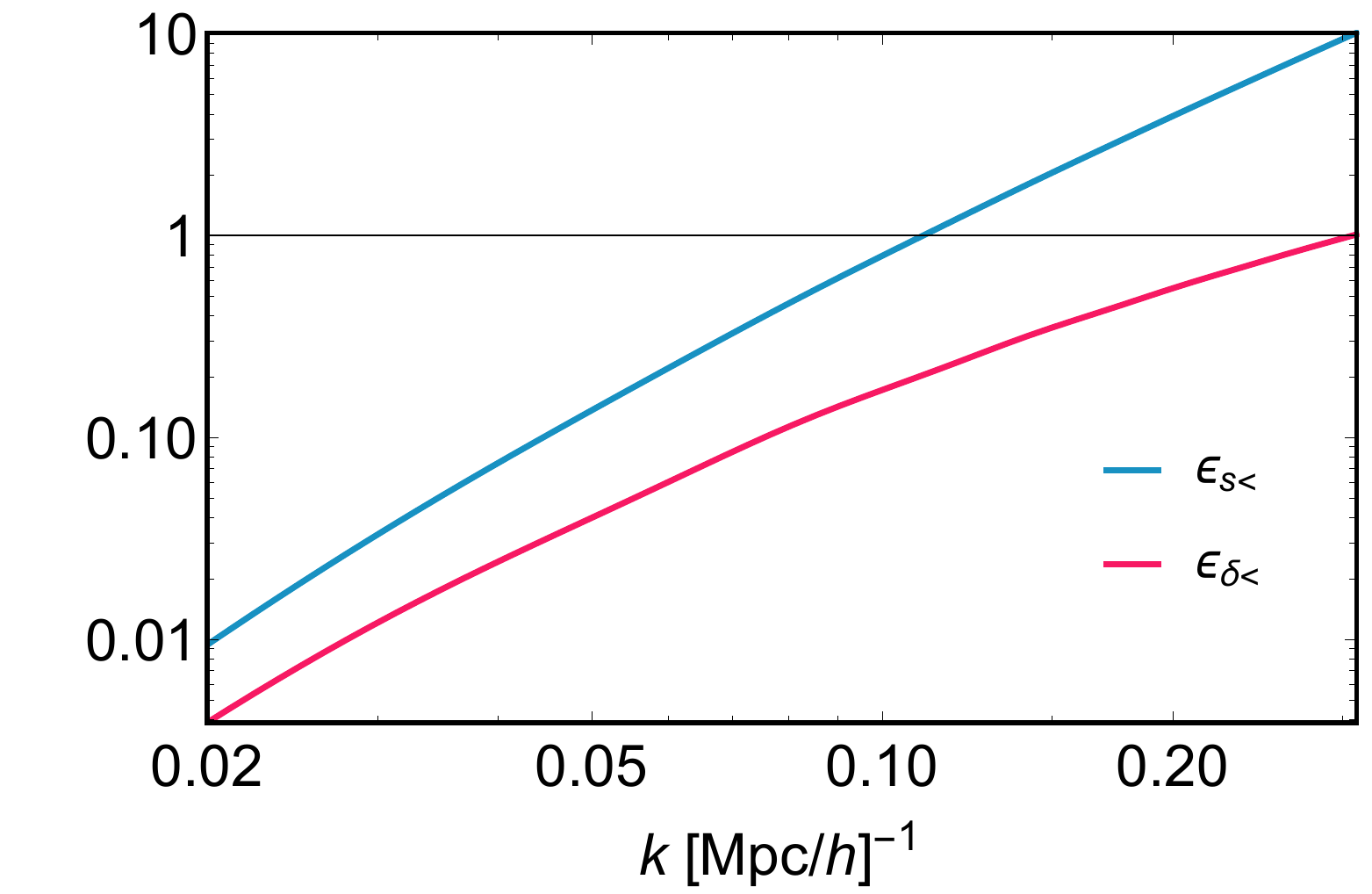} 
\caption{ We plot the expansion parameters of Eulerian perturbation theory, $\edl ( k)$ which describes the effects of IR density fluctuations, and $\esl(k)$ which describes the effects of IR displacements.  We see that IR displacements are significantly larger than the IR density fluctuations.     } \label{epsilonsfig}
\end{figure*}

As written in \eqn{Aalphadef}, \eqn{Balphadef}, and \eqn{k3def}, it appears that $A_\alpha^2 - B_\alpha \sim  \mathcal{O}(\lambda_D \lambda_{\Delta v})$ (we will momentarily show that this is in fact not true, but it is instructional to consider for now), since $\Delta A_{\alpha}$ and $\Delta B_\alpha \equiv B_\alpha -1$ contain terms linear in $\lambda_D \lambda_{\Delta v}$.  This appears to be the case because, for example, one can insert a $\Lambda$CDM solution into a vertex proportional to $\Delta v^i$ in \eqn{IReom}, so that the whole vertex is only proportional to one power of $\lambda_{\Delta v}$.  Continuing in this way to higher loops, one would obtain  
\be \label{nloopir}
P_{N\text{-loop}}^{\rm IR} ( k ) \sim P_{11}(k) \, \lambda_D \lambda_{\Delta v} \,  \esl(k)^N \ . 
\ee
This should to be contrasted with the situation in $\Lambda$CDM, where the higher order corrections are, roughly, 
\be
P_{N\text{-loop}}^{\Lambda\text{CDM}} ( k ) \sim P_{11}(k) \, \edl ( k )^N  \ . 
\ee
It is important to note that the corrections in \eqn{nloopir} are generically proportional to $\lambda_{\Delta v}$, and not $\lambda_{\Delta v}^N$, so that it really is $\esl$ that controls the expansion, and not $\lambda_{\Delta v} \, \esl$.  What this means is that even if $\lambda_{\Delta v} \, \esl (k ) \ll 1$, corrections higher than one loop will contribute powers of $\esl ( k )$, and so one must be in a regime where $\esl ( k ) \lesssim 1$ for the perturbative series to be sensible.  Looking at \figref{epsilonsfig}, this would mean that we would only be able to trust our perturbative calculations well below $k \approx 0.1 \unitsk$.  

Luckily, this is not the case, and we will show that in general, for small $\lambda_D$,
\be \label{nloopirtrue}
P_{N\text{-loop}}^{\rm IR} ( k ) \sim P_{11}(k) \,  \left[ \lambda_D \lambda_{\Delta v}^2 \,  \esl(k) \right]^N \ .
\ee
To see this, we will solve \eqn{IReom} in a more illuminating way, by first changing coordinates to the frame comoving with the dark matter, perturbatively solving the equations in that frame, and then changing coordinates back.  Consider the coordinate transformation (note that we are \emph{not} also shifting the fields),
\be \label{dmcoordchange}
\tilde x^i = x^i + n_L^i ( \xvec , t ) \ , \quad \delta ( x^i , t ) = \tilde \delta ( \tilde x^i , t) \, \quad \text{and} \quad \dot n_L^i ( \xvec , t) = - a^{-1} v^i_{(1)} ( \xvec , t)  \ ,
\ee
so that in momentum space, the fields are related by
\be \label{coordtransf2tilde}
\delta_{\kvec}  = \int d^3 x \int_{\kvec'} e^{i ( \kvec'- \kvec )\cdot \xvec}  \exp \left\{ \int_{\qvec} \frac{\qvec \cdot \kvec'}{q^2} \delta^{(1)}_{\qvec} e^{i \qvec \cdot \xvec}  \right\} \tilde \delta_{\kvec'} \ .
\ee
{For the two-point function, this gives\footnote{{Notice that, since our change of coordinates does not break translation invariance or isotropy, we still have the general form $\langle \tilde \delta_{\kvec} \tilde \delta_{\kvec'} \rangle = ( 2 \pi)^3 \delta_D ( \kvec+ \kvec') \tilde P ( k )$.  This is to be contrasted with, for example, redshift space distortions, which are based on a similar change of coordinates, namely $\xvec \rightarrow \xvec -  \hat x [ \vec{v} ( \xvec , t) \cdot \hat x] /(a H)$.  Since this transformation breaks translation invariance by defining a center point, the power spectrum is a general function of $\kvec$ and $\kvec'$ \cite{Hamilton:1995px, Zaroubi:1993qt}.  In the more commonly used plane-parallel approximation, redshift space distortions are based on the transformation $\xvec \rightarrow \xvec -  \hat z [ \vec{v} ( \xvec , t) \cdot \hat z] /(a H)$ which does not break translations, but does break isotropy.  Thus the power spectrum in that case is diagonal in $\kvec$ and $\kvec'$, but depends on $\kvec$ (as opposed to just $k$).}} 
\be \label{resum1}
\langle \delta_{\kvec_1} \delta_{\kvec_2}  \rangle  = \left( \prod_{a=1}^2 \int d^3 x_a \int_{\kvec_a'} e^{i \xvec_a \cdot ( \kvec_a' - \kvec_a)} \right) \left\langle \exp \left\{ \sum_{b=1}^2 \int_{\qvec} \frac{\qvec \cdot \kvec_b'}{q^2} \delta_{\qvec}^{(1)} e^{i \qvec \cdot \xvec_b} \right\}  \tilde \delta_{\kvec_1'} \tilde \delta_{\kvec_2'}  \right\rangle \ .
\ee
To find the expression at leading order in $k / q$, we can expand $e^{i \qvec \cdot \xvec_b} \approx 1 + i \qvec \cdot \xvec_b + \dots $  inside of the exponential, and then expand the exponential itself.  At each order in $\delta$, we want the most factors of $k/q$ (and therefore the least number of factors of $q$ in the numerator), so we see that we can actually set $e^{i \qvec \cdot \xvec_b}\rightarrow 1$ to compute the leading contribution.  In this case, the integrals over $d^3 x_1$ and $d^3 x_2$ are trivial, and one ends up with  }
\be \label{coordchps}
\langle \delta_{\kvec_1} \delta_{\kvec_2} \rangle \approx \langle \tilde \delta_{\kvec_1} \tilde \delta_{\kvec_2} \rangle \ . 
\ee

The equation of motion for the tilde field evaluated at the tilde coordinates is given by 
\be \label{aftercoord}
\ddot{\tilde \delta} + \bar \nu_\Phi \dot{\tilde \delta} - \bar \mu_{\Phi} \tilde \delta \approx   \frac{-a^{-1}}{1 - \sigma_\Phi} \Big( \nu_\Phi \Delta v^i \partial_i \tilde \delta + \sigma_\Phi \left( 2 \Delta v^i \partial_i \dot{\tilde \delta} + \Delta \dot v^i \partial_i \tilde \delta - H \Delta v^i \partial_i \tilde \delta \right) - \sigma_\Phi a^{-1} \Delta v^i \Delta v^j \partial_i \partial_j \tilde \delta   \Big) \ .
\ee
Now, since we are in the frame of the dark matter, the only velocity that appears is the relative velocity, as expected, and it is now clear that each perturbative order gets an extra power of $\lambda_{\Delta v}$.  The linear solutions are the same, $\tilde \delta^{(1)}(t) = \delta^{(1)} ( t ) $, and the second order solution is given by 
\be \label{deltatwoleadingir}
\tilde \delta^{(2)} ( t ) \approx \Delta \tilde A_{\alpha} ( t )   \frac{\partial_i \delta^{(1)} ( t ) }{\partial^2} \partial_i \delta^{(1)} ( t ) \ , 
\ee
where 
\be
\Delta \tilde A_{\alpha} ( t ) = \Delta A_{\alpha} ( t ) \ ,
\ee
which is given in \eqn{Aalphadef}.  Finally, the third order solution is given by 
\be \label{deltathreeleadingir}
\tilde \delta^{(3)} ( t ) \approx \Delta \tilde B_{\alpha} ( t )    \half   \frac{ \partial_i  \delta^{(1)} ( t)}{\partial^2}   \frac{ \partial_j  \delta^{(1)} ( t)}{\partial^2}  \partial_i \partial_j   \delta^{(1)} ( t)  \ , 
\ee
where
\be \label{tildebalphafn}
\Delta \tilde B_\alpha ( t ) = \int_0^t d t_1 \, \bar G( t , t_1 ) \tilde K_3 ( t_1 ) \frac{D_+ ( t_1)^3}{D_+(t)^3} \ ,
\ee
and
\be \label{tildek3exp}
\tilde K_3 = \frac{2}{1 - \sigma_\Phi} \left( \nu_\Phi L_{\Delta v } \Delta \tilde A_\alpha + \sigma_{\Phi} \left( L_{\Delta v} (L_{\Delta v} + 2 \Delta \dot{\tilde A}_\alpha + 5 f H \Delta \tilde A_\alpha  ) + \dot L_{\Delta v} \Delta \tilde A_\alpha   \right)   \right)   \ . 
\ee
One can then change coordinates back to the non-tilde fields using \eqn{coordtransf2tilde} in order to relate $\Delta \tilde B_\alpha$ to $\Delta A_\alpha$ and $\Delta B_\alpha$, but as we show next, we can do this by simply matching the expressions for $P^{\rm IR}_{1\text{-loop}}$ using \eqn{coordchps}.

\begin{figure*}[t] 
\centering 
\hspace{-.3in} \includegraphics[width=.63\textwidth]{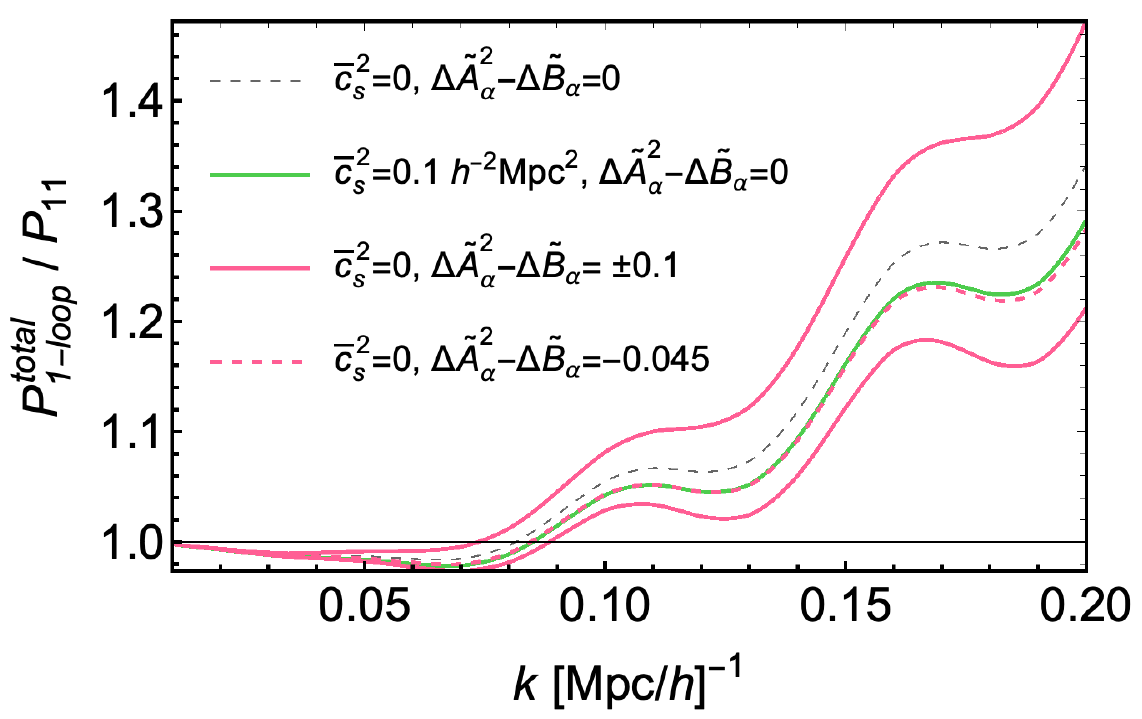} 
\caption{ We plot the effect on the one-loop power spectrum from the IR contribution \eqn{p1looplimitcoordch}, which arises from the violation of the consistency relations in DHOST theories.  Specifically, we plot \eqn{fullp1loop} where $P_{11}$ and $P_{1\text{-loop}}$ are computed in $\Lambda$CDM, and $P_{1\text{-loop}}^{\rm IR}$ is given in \eqn{p1looplimitcoordch}.  We see that the effect of the LSS counterterm is similar to the IR contribution \eqn{p1looplimitcoordch}, i.e. both are proportional to $k^2 P_{11} ( k )$.  In particular, we see that the effect of the IR contribution with $\Delta \tilde A_\alpha^2 - \Delta \tilde B_\alpha = - 0. 045$ is very similar to the effect of the counterterm with $\bar c_s^2 = 0.1 (\unitsr)^{2}$, which is a typical value (see for example \cite{Foreman:2015lca}).} \label{p1loops}
\end{figure*}

Computing the power spectrum with \eqn{coordchps} and the expressions above, we obtain
\be \label{p1looplimitcoordch}
 P_{1\text{-loop}}^{\rm IR} (k ) =  (\Delta \tilde A_\alpha^2 -  \Delta \tilde B_\alpha ) P_{11} ( k ) \int_{q \lesssim k}  \frac{d^3 q}{(2 \pi)^3} \left( \frac{\qvec \cdot \kvec}{q^2} \right)^2 P_{11} ( q )  \  .
\ee 
Of course, this has to be equal to the expression in \eqn{p1looplimit}, so we see that we must have 
\be \label{deltabeq}
\Delta B_\alpha = 2 \Delta A_\alpha + \Delta \tilde B_\alpha \ ,
\ee
which is the promised simplification of the solution \eqn{Balphadef}.  Looking at the expressions in \eqn{Aalphadef} and \eqn{tildek3exp}, we see that $\Delta A_\alpha \sim \mathcal{O}(\lambda_D \lambda_{\Delta v})$, and that $\Delta \tilde B_{\alpha} \sim \mathcal{O}( \lambda_D^2 \lambda_{\Delta v}^2) + \mathcal{O}( \lambda_D \lambda_{\Delta v}^2)$.  
This means that, for small $\lambda_D$,
\be \label{1loopir2}
P_{1\text{-loop}}^{\rm IR} ( k ) \sim P_{11}(k) \,  \lambda_D  \lambda_{\Delta v}^2 \,  \esl(k) \ ,
\ee
which is proportional to the power spectrum of the relative displacements (given by $\lambda_{\Delta v}^2 \esl$).  Indeed as we continue in the loop expansion, we will obtain \eqn{nloopirtrue}, showing that the perturbative expansion is in powers of the power spectrum of the relative displacements.  While one could have shown \eqn{deltabeq} directly from \eqn{Balphadef} by using properties of the Green's function, it is in practice quite non-trivial.\footnote{With the benefit of the simplified solution \eqn{deltabeq}, we were able to check it directly.  To do so, start with \eqn{k3def}, replace $f = \dot D_+ / ( H {D_+})$, and plug in the definition of $\Delta A_\alpha$ \eqn{Aalphadef}.  Then, use the identity
\be
\bar G( t , t_2 ) D_+ ( t_1 ) = \bar G( t_1 , t_2 )  \left( D_+ ( t ) - \bar G ( t , t_1 ) \dot D_+ ( t_1) \right)  + \bar G( t , t_1 ) D_+ ( t_1) \partial_{t_1} \bar G( t_1 , t_2 ) \ .
\ee 
Finally, use the fact that 
\be
\int^t d t_1 \dot D_+ ( t_1 ) \int^{t_1} d t_2 \, g ( t , t_2 ) = D_+(t)  \int^t d t_1  \, g ( t , t_1) - \int^t d t_1 D_+ ( t_1 ) g ( t , t_1)\ ,
\ee
for a generic function $g(t,t_1)$.  }

In \figref{p1loops}, we show the effect of the one-loop IR contribution \eqn{p1looplimitcoordch} which arises in DHOST theories due to the violation of the consistency relations.  In particular, we plot
\be \label{fullp1loop}
P^{\rm total}_{1\text{-loop}} (k) =  P_{1\text{-loop}} ( k ) - 2 (2 \pi) \bar c_s^2 k^2 P_{11}(k) + P_{1\text{-loop}}^{\rm IR} ( k ) \ , 
\ee
where we have included the one-loop counterterm contribution, proportional to $\bar c_s^2$, from the effective field theory of large-scale structure \cite{Baumann:2010tm, Carrasco:2012cv}.  As we can see from \eqn{p1looplimitcoordch}, the effect of the IR contribution is of the same form, proportional to $k^2 P_{11}(k)$, as the EFT counterterm.  Although these effects are the same in the one-loop power spectrum (unfortunately making it difficult to measure the violation of the consistency relations in this way), we stress that the origins of these terms are very different.  The EFT counterterm arises from uncomputable UV physics, and the coefficient $\bar c_s^2$ is a free parameter which should be matched to data in order to consistently incorporate the UV in the perturbative calculation. Different values of $\bar c_s^2$ do \emph{not} correspond to different IR theories.  The IR contribution, on the other hand, is computable within the theory, since it only depends on IR modes: different values of $\Delta \tilde A_\alpha^2 - \Delta \tilde B_\alpha$ correspond to physically different IR theories.  That said, we can place a loose bound on the size of the IR contribution by saying that it should not be much larger than the typical size of the counterterms measured from galaxy clustering data \cite{DAmico:2019fhj, Ivanov:2019pdj}, which is approximately $\bar c_s^2 \approx 0.1 (\unitsr)^2$.  From \figref{p1loops}, we see that this means that, assuming no large cancellations between contributions,
\be \label{roughconstraint}
|\Delta \tilde A_{\alpha}^2 - \Delta \tilde B_\alpha | \lesssim 0.045 \ .
\ee
After assuming constraints imposed by avoiding gravitational wave decay \cite{Creminelli:2018xsv, Creminelli:2019nok} and instabilities induced by gravitational waves \cite{Creminelli:2019kjy}, \eqn{roughconstraint} is essentially constraining the size of $\beta_1$.  However, connecting this to a rigid constraint on the DHOST parameters would require a fairly involved analysis, so we leave a detailed study for future work.

As our next example, we look at the tree-level trispectrum.  With the solution for $\delta^{(3)}$, we can also compute the violation of the consistency relations in the double soft limit of the four-point function.  We obtain, after setting $\kvec_1 \approx - \kvec_2$, 
\be \label{trispectrumlimit}
\lim_{q_1 , q_2 \rightarrow 0} \frac{T_t ( \qvec_1 , \qvec_2 , \kvec_1 , \kvec_2 )}{P_{11} ( q_1 ) P_{11} ( q_2 ) P_{11}( k_1) } \approx - 2 ( \Delta \tilde A_\alpha^2 - \Delta \tilde B_\alpha) \frac{\qvec_1 \cdot \kvec_1}{q_1^2} \frac{\qvec_2 \cdot \kvec_1}{q_2^2} \ .
\ee
For theories that satisfy the consistency relations, \eqn{doublesoft} says that this term is zero.  However, we see that because the different diagrams have different coefficients, the cancellation no longer happens in DHOST theories.  This result is intimately connected to the one-loop power spectrum result, and the two are essentially two sides of the same coin: the IR contribution in the one-loop power spectrum can be computed by connecting the two soft legs of the trispectrum with a linear power spectrum and integrating over the intermediate momentum.

%
%
\subsection{IR-resummation} \label{irresumsec}

Finally, we discuss how the violation of the consistency relations affects the evolution of the BAO signal in the matter power spectrum.  In $\Lambda$CDM, the consistency relations are satisfied and one can resum the effects of the long-wavelength velocity on the BAO signal.  The reasons for this are similar to those presented in \secref{bispsubsec}, namely, that the leading IR contributions are fixed by the effect of the coordinate transformation \eqn{coordtransf2}.  Even though the leading IR contributions cancel in the extreme IR limit of the equal-time power spectrum (see \eqn{irloops}), the effects on oscillatory features can be large and are computable with \eqn{coordtransf2}, as in \eqn{bispecslsubosc} for the bispectrum.  This procedure is called the IR-resummation and has been considered for a single species in $\Lambda$CDM in, for example,  \cite{Senatore:2014via, Matsubara:2007wj, Lewandowski:2015ziq, Senatore:2017pbn, Lewandowski:2018ywf, Baldauf:2015xfa, Vlah:2015sea, Vlah:2015zda, Blas:2016sfa, delaBella:2017qjy, Ivanov:2018gjr}.

The situation is different for theories with a large-scale relative velocity which cannot be eliminated by a coordinate transformation, though.  This includes the DHOST theories studied in this paper, and also theories with multiple species, such as dark matter and baryons (see \cite{Bernardeau:2012aq} for a treatment using the eikonal approximation, and \cite{Lewandowski:2014rca} using the IR-resummation).  Basically, as we will show below, one can boost to the frame of one of the species, and that large-scale velocity can be resummed.  However, there are perturbative effects of the large-scale relative velocity which cannot be resummed, and so must continue to be treated perturbatively.  

In this work, we focus on the DHOST theories described in \secref{dhostsec}, although we expect our results to be straightforwardly applicable to the two-fluid case.  Since we cannot eliminate both velocities with a single coordinate transformation, we have to choose one to resum.  For concreteness, we choose the dark-matter velocity and consider the coordinate transformation defined in \eqn{dmcoordchange}.  In these coordinates, the equation of motion for $\tilde \delta$ is given by \eqn{aftercoord}, plus a series of other non-linear terms coming from the full DHOST theory (i.e. the terms that are not the leading IR terms and therefore we did not consider in this work) and from the coordinate transformation, which can be expanded to the desired order in perturbation theory.  Then, the power spectrum in the original, non-tilde coordinates is given by \eqn{resum1}, where the $\tilde \delta$ fields can be expanded perturbatively, and the large-scale dark-matter velocity is resummed in the exponential.  {In this section, we consider the effects only of soft loops, we will work explicitly up to one loop in the relative velocity, and we will resum to all orders the leading IR terms of the dark-matter velocity.}

 To organize our calculation, we can write the solutions in the new coordinate system as
\begin{align} \label{irexpandfordelta}
\tilde \delta_{\kvec}^{(n)}  \equiv \tilde \delta_{{\rm IR} , \kvec}^{(n)}  + \Delta \tilde \delta_{ \kvec}^{(n)}  \ , 
\end{align}
where $\tilde \delta^{(n)}_{{\rm IR}, \kvec} \sim ( k/q)^{n-1} [  \lambda_{\Delta v}\, \delta^{(1)}_{\qvec}]^{n-1} \delta^{(1)}_{\kvec}$ for $n \geq 2$ is the leading IR term (given by \eqn{deltatwoleadingir} and \eqn{deltathreeleadingir} for $n=2$ and $n=3$ respectively), and $\Delta \tilde \delta_{ \kvec}^{(n)} $ is the rest of the solution.  In particular, we have $\tilde \delta^{(1)}_{{\rm IR},\kvec}  = 0$ and $\Delta \tilde \delta_{ \kvec}^{(1)}  = \delta^{(1)}_{{ \kvec}} $.  For $\Lambda$CDM, where one can eliminate all of the leading IR terms with a coordinate transformation, $ \tilde \delta_{{\rm IR} , \kvec}^{(n)}  = 0$ for all $n$.  

  At this point, it is instructive to see how \eqn{resum1} gives the standard IR-resummation in $\Lambda$CDM for a single fluid.  Since in $\Lambda$CDM we have $ \tilde \delta_{{\rm IR} , \kvec}^{(n)}  = 0$, the tilde fields are not correlated with the IR field in the exponent, so that one has \cite{Creminelli:2013poa}
\be \label{nolongmodes}
 \left\langle \exp \left\{ \sum_{b=1}^2 \int_{\qvec} \frac{\qvec \cdot \kvec_b'}{q^2} \delta_{\qvec}^{(1)} e^{i \qvec \cdot \xvec_b} \right\}  \tilde \delta_{\kvec_1'} \tilde \delta_{\kvec_2'}  \right\rangle  {\approx}  \left\langle \exp \left\{ \sum_{b=1}^2 \int_{\qvec} \frac{\qvec \cdot \kvec_b'}{q^2} \delta_{\qvec}^{(1)} e^{i \qvec \cdot \xvec_b} \right\}  \right\rangle \langle  \tilde \delta_{\kvec_1'} \tilde \delta_{\kvec_2'}  \rangle  \ .
\ee
Then, using the cumulant theorem for a Gaussian field $X$ centered around zero, $\langle e^X \rangle = e^{\half \langle X^2 \rangle}$, changing variables of integration in \eqn{resum1} to $\rvec = \xvec_1 - \xvec_2$ and $\xvec = \half ( \xvec_1 + \xvec_2)$,  integrating over $d^3 x$, and finally integrating over $d^3 k_2'$, we find the IR-resummed power spectrum
\be \label{lcdmresum}
P^{\Lambda}(k)  {\approx} \int d^3 r \int_{\kvec'} e^{i \rvec \cdot (\kvec' - \kvec)} \exp \left\{- \half \int_{\qvec }P_{11}( q )  \left( \frac{\qvec \cdot \kvec'}{q^2} \right)^2  \left( 2 - e^{i \qvec \cdot \rvec} - e^{- i \qvec \cdot \rvec} \right)    \right\}  \tilde P^{\Lambda} ( k')  \ ,
\ee
where we have used the superscript $\Lambda$ as a reminder that this is valid in $\Lambda$CDM with a single fluid. {Furthermore, since we are interested in resumming the leading IR contributions of only soft loops at each order, we can take $\tilde P^\Lambda ( k' ) \rightarrow P_{11} ( k')$.}\footnote{{To include higher loops in the Eulerian expansion, one can follow \cite{Senatore:2014via}.}}

In DHOST theories, \eqn{nolongmodes} is no longer valid because the tilde fields still depend on the long modes through the relative velocity.  To deal with this, we expand the tilde fields on the left-hand side of \eqn{nolongmodes} as in \eqn{irexpandfordelta}.  For notational convenience, we define
\be
E[\delta^{(1)}  ] \equiv \exp \left\{ \sum_{b=1}^2 \int_{\qvec} \frac{\qvec \cdot \kvec_b'}{q^2} \delta_{\qvec}^{(1)} e^{i \qvec \cdot \xvec_b} \right\}  \  . 
\ee
Then, expanding the tilde fields on the left-hand side of \eqn{nolongmodes}, we have
\begin{align} \label{newresumexpand1}
\left\langle E[\delta^{(1)}  ] \, \tilde \delta_{\kvec_1 '} \tilde \delta_{\kvec_2'} \right\rangle  = &  \left\langle E[\delta^{(1)} ] \, \Delta \tilde \delta_{\kvec_1 '} \Delta \tilde  \delta_{\kvec_2'} \right\rangle   \\
& +  \left\langle E[\delta^{(1)} ] \, \left( \Delta \tilde \delta_{\kvec_1 '}^{(1)}  \tilde  \delta_{{\rm IR},\kvec_2'}^{(2)} +  \tilde  \delta_{{\rm IR},\kvec_1'}^{(2)}  \Delta \tilde \delta_{\kvec_2 '}^{(1)}   \right)  \right\rangle   \nonumber \\
& +  \left\langle E[\delta^{(1)}  ] \, \left( \Delta \tilde \delta_{\kvec_1 '}^{(1)}  \tilde  \delta_{{\rm IR},\kvec_2'}^{(3)} +  \tilde  \delta_{{\rm IR},\kvec_1'}^{(3)}  \Delta \tilde \delta_{\kvec_2 '}^{(1)}  + \tilde \delta^{(2)}_{{\rm IR},\kvec_1'}  \tilde \delta^{(2)}_{{\rm IR},\kvec_2'} \right)  \right\rangle  + \dots \ , \nonumber
\end{align}
where we have used $\Delta \tilde \delta_{\kvec} \equiv \sum_n \Delta \tilde \delta_{\kvec}^{(n)}$.  We have organized this expansion in powers of $\lambda_{\Delta v}$, with the first line being proportional to $\lambda_{\Delta v}^0$, the second line proportional to $\lambda_{\Delta v}^1$, the third line proportional to $\lambda_{\Delta v}^2$ (these are all of the terms necessary for the one-loop calculation), and the $\dots$ standing for terms higher order in $\lambda_{\Delta v}$.  By keeping $E$ in its exponentiated form, we resum all of the leading IR terms from the dark-matter velocity for each power of $\lambda_{\Delta v}$.  For example, the first line resums all of the terms of the form $\lambda_{\Delta v}^0 (k/q)^{2n} P_{11}(q)^{n} P_{11}(k)$ for $n\geq0$, the second line resums all of the terms of the form $\lambda_{\Delta v}^1 (k/q)^{2n} P_{11} ( q)^n P_{11} ( k)$ for $n \geq1$, and the third line resums all of the terms of the form $\lambda_{\Delta v}^2 (k/q)^{2n} P_{11} ( q)^n P_{11} ( k)$ for $n \geq1$. 

Defining the power spectrum of the terms with no long modes as 
\be
\langle \Delta \tilde \delta_{\kvec} \Delta \tilde \delta_{\kvec'} \rangle = ( 2 \pi)^3 \delta_D ( \kvec + \kvec' ) P_{\Delta \tilde \delta } ( k ) \ , 
\ee
 plugging \eqn{newresumexpand1} into \eqn{resum1}, and using properties of the correlation functions involving $E$ given in \appref{resumapp}, we find that the power spectrum with the long-wavelength dark-matter velocity resummed, up to order $\lambda_{\Delta v}^2$, is 
\begin{align}
P (  k ) {\approx} &  \int d^3 r \int_{\kvec '} e^{i \rvec \cdot ( \kvec' - \kvec)} \exp \left\{ -\half \int_{\qvec_3} \left( \frac{\qvec_3 \cdot \kvec'}{q_3^2}\right)^2   P_{11}(q_3) \left( 2 - e^{i \qvec_3 \cdot \rvec} - e^{-i \qvec_3 \cdot \rvec} \right) \right\} \Bigg[ P_{\Delta \tilde \delta} ( k') \nonumber \\
& +   \int_{\qvec} \left( \frac{\qvec \cdot \kvec'}{q^2}\right)^2 P_{11} ( q )  \left( 2 \Delta A_\alpha \left(  e^{- i \qvec \cdot \rvec} - 1 \right) P_{11} ( k') -\Delta \tilde B_\alpha P_{11} ( k') + \Delta A_\alpha^2 P_{11}( | \kvec' - \qvec| )  \right) \nonumber \\
& + \int_{\qvec_1 , \qvec_2} \left( \frac{\qvec_1 \cdot \kvec'}{q_1^2} \frac{\qvec_2 \cdot \kvec'}{q_2^2}   \right)^2 P_{11} ( q_1) P_{11} ( q_2) \left( 1 - e^{-i \qvec_1 \cdot \rvec}  - e^{-i \qvec_2 \cdot \rvec}  + e^{-i (\qvec_1 + \qvec_2) \cdot \rvec}  \right) \nonumber \\
& \hspace{1in} \times \left( \Delta \tilde B_\alpha P_{11} ( k') - \Delta A_\alpha^2 P_{11} ( | \kvec' - \qvec_1| )  \right) + \mathcal{O}(\lambda_{\Delta v}^3 )\Bigg] \ .  \label{irresumnew}
\end{align}
{Again, to the order that we work, we can take $P_{\Delta \tilde \delta} ( k') \rightarrow P_{11} ( k')$, and } the first line above is of the same form as the single-fluid $\Lambda$CDM resummation \eqn{lcdmresum}.  This is because, after the coordinate transformation, $P_{\Delta \tilde \delta}$ does not contain any long modes, which is analogous to what happens in $\Lambda$CDM for the full power spectrum $\tilde P^\Lambda$.  However, the remaining lines above are perturbative corrections proportional to powers of the large-scale relative velocity.  In all terms, though, the large-scale dark matter velocity has been fully resummed.  

As a check, we can compare the above to the one-loop result from perturbation theory.  Using \eqn{delta2solde} and \eqn{delta3sol} , we have 
\be \label{p1looplimit2}
P_{1\text{-loop}} ( k ) \approx    \int_{q \lesssim k}  \frac{d^3 q}{(2 \pi)^3} \left( \frac{\qvec \cdot \kvec}{q^2} \right)^2 P_{11} ( q )  \left( A_\alpha^2 P_{11} ( | \kvec - \qvec|) - B_\alpha P_{11} ( k ) \right) \ .
\ee 
Then using $A_\alpha = 1 + \Delta A_\alpha$ and $B_\alpha = 1 + 2 \Delta A_\alpha + \Delta \tilde B_\alpha$, we see that this exactly matches the one-loop, leading IR perturbative piece of \eqn{irresumnew}.  We leave a full exploration of the quantitative effects of \eqn{irresumnew} on the BAO for future work.

%
%
%
\section{Conclusions} \label{conclusionssec}

In this work, we have investigated the consequences of the violation of the consistency relations in theories with dark energy.  The gravitational field equations for DHOST theories \eqns{phieq1}{pieq1} contain terms with time derivatives and single spatial derivatives.  The structure of these terms is fixed by the symmetry \eqn{transf1}, and in particular, the transformation of $\pi$ is fixed by the associated coordinate change, see \eqn{bipi}.  This means that in the combined fluid and gravitational field equations \eqn{IReom}, although invariant under an overall Galilean transformation, the long-wavelength effects of both $v^i$ and $\partial_i \pi$ cannot be eliminated with a single coordinate transformation (i.e. one cannot construct a physical adiabatic mode), contrary to the case in Horndeski theories.  Thus, the standard $\Lambda$CDM consistency relations are violated by terms proportional to the large-scale relative velocity $\Delta v^i \equiv v^i + a^{-1} \partial_i \pi$ (see also \cite{Crisostomi:2019vhj}).  

Although the consistency relations are violated, we show that the leading effects in the IR are determined by the linear equations and the symmetries of the fluid system.  In this way, we have computed the effects of the violation on the BAO, the bias expansion, the one-loop power spectrum, and the tree-level trispectrum, explicitly showing how these effects depend on the relative velocity $\Delta v^i$.  Specifically, in \secref{bispsubsec}, we showed that the size of the BAO oscillations in the squeezed limit of the bispectrum can be modified.  In \secref{biassec} we showed how the bias expansion contains terms proportional to the relative velocity which contribute to the squeezed limit of the galaxy bispectrum.  In \secref{powerspectrumsec}, we computed the contribution from IR modes to the one-loop power spectrum and discussed how the perturbative expansion is altered and can depend more strongly on the IR displacements.  {We also computed the effect of the violation of the consistency relations on the double soft limit of the tree-level trispectrum.  Finally, in \secref{irresumsec}, we showed how one can resum the effects of \emph{one} of the large-scale velocities in the power spectrum to correctly describe the BAO. After doing that, the effect of the relative velocity must be included perturbatively.   Our IR-resummation results are also applicable to other systems with a large-scale relative velocity, for example, dark matter and baryons.}

The effects that we have computed are proportional to two separate parameters, $\lambda_D$ and $\lambda_{\Delta v}$.  The parameter $\lambda_D$ stands for the DHOST parameters, $\alphaH$ and $\beta_1$ (which enter the linear equation for $\delta$ through $\nu_\Phi / H$ and $\sigma_\Phi$), and $\lambda_{\Delta v} \sim L_{\Delta v} / H$ measures the relative velocity.  Both of these parameters have to be non-zero to see violations of the consistency relations in the ways that we have discussed.  Horndeski theories, for example, have $\lambda_{\Delta v} \sim \mathcal{O}(1)$ but $\lambda_D = 0$, and \emph{do} satisfy the consistency relations \cite{Crisostomi:2019vhj}.  Similarly, we can have DHOST theories with $\lambda_D \neq 0$, but if $\lambda_{\Delta v } = 0$, they will still satisfy the consistency relations as well.

In our calculations, because there is not generally an EdS-type approximation for the time dependence in modified gravity theories, we have considered the exact time dependence of the linear equations, i.e. we have used the full Green's function.  Thus, our results are also relevant for perturbation theory in $\Lambda$CDM with exact time dependence.

The parameter space of dark-energy theories, including DHOST theories, is already well constrained by observations dealing with, for example, properties of gravitational-wave propagation, instabilities in the dark energy sector induced by gravitational waves, and Vainshtein screening.  However, it could be the case that the UV cutoff of the EFT of dark energy is much smaller than previously thought \cite{PhysRevLett.123.251103}, in which case some of the constraints placed on the EFT parameters from small scales (for example in the Vainshtein regime) may not be valid.  If this is the case, LSS effects, such as the ones we have discussed in this work, will be more relevant because they happen well within the regime of validity of the EFT of dark energy.  We leave a detailed study of the parameter constraints for future work.

%
\section*{Acknowledgements}

The author would like to thank Leonardo Senatore and Filippo Vernizzi for helpful correspondence and discussion.  M.~L.~acknowledges financial support from the European Research Council under ERC-STG-639729, \emph{preQFT: Strategic Predictions for Quantum Field Theories}.   

%
%

\appendix

%
%
%
\section{Review of derivations of the consistency relations} \label{crreviewapp}

In this Appendix, we present one method of deriving the consistency relations in $\Lambda$CDM, called the background wave argument, and we mostly follow \cite{Creminelli:2013poa}.  First, we would like to compute the leading terms of 
\be
\langle \delta_{\qvec_1} ( \tau_1 ) \delta_{\kvec_1} ( t_1 ) \cdots \delta_{\kvec_n} (t_n)\rangle 
\ee
and 
\be
\langle \delta_{\qvec_1} ( \tau_1 ) \delta_{\qvec_2} ( \tau_2 ) \delta_{\kvec_1} ( t_1 ) \cdots \delta_{\kvec_n} (t_n)\rangle 
\ee
in the limit $q_i \ll k_j$.  Using the definition of conditional probability, we have the following expressions
\begin{align}
\begin{split} \label{backgroundvel}
 \langle \delta_{\qvec_1} ( \tau_1 ) \delta_{\kvec_1} ( t_1 ) \cdots \delta_{\kvec_n} (t_n)\rangle & = \langle \delta_{\qvec_1} ( \tau_1 ) \langle   \delta_{\kvec_1} ( t_1 ) \cdots \delta_{\kvec_n} (t_n)\rangle_{v^i_L} \rangle \ , \\
 \langle \delta_{\qvec_1} ( \tau_1 )  \delta_{\qvec_2} ( \tau_2 )  \delta_{\kvec_1} ( t_1 ) \cdots \delta_{\kvec_n} (t_n)\rangle & = \langle \delta_{\qvec_1} ( \tau_1 ) \delta_{\qvec_2} ( \tau_2 ) \langle   \delta_{\kvec_1} ( t_1 ) \cdots \delta_{\kvec_n} (t_n)\rangle_{v^i_L} \rangle \ ,
\end{split}
\end{align}
where the notation $\langle \, \cdot \, \rangle_{v^i_L}$ means that we only average over configurations which at some fixed time have the large-scale velocity $v^i_L$.  
From the expressions \eqn{backgroundvel}, we see that we first need to compute the correlation function of the short modes given the large-scale velocity.  

The symmetry \eqn{transf2} tells us that 
\be \label{noxsym}
\langle \delta ( x^i_1 , t_1) \cdots \delta ( x^i_n , t_n) \rangle_{v^i_L} = \langle  \delta ( \tilde x^i_1 , t_1 ) \cdots  \delta ( \tilde x^i_n , t_n) \rangle_{v^i_L + a \dot n^i }  
\ee
so that we can relate a correlation function with a large-scale velocity $v^i_L$ to a correlation function with a large-scale velocity $v^i_L + a \dot n^i$.  This is only strictly true if $n^i$ has no spatial dependence, so that we only change the zero mode of the background velocity.  However, if we give $n^i$ a weak spatial dependence, \eqn{noxsym} will capture the leading effects, but of course will be corrected by higher derivative terms. Thus, we have
\be \label{realcoordchg}
\langle \delta ( x^i_1 , t_1) \cdots \delta ( x^i_n , t_n) \rangle_{v^i_L}   \approx \langle  \delta ( x^i_1 +  n^i_L(\xvec_1 , t_1) , t_1 ) \cdots  \delta ( x^i_n +  n^i_L ( \xvec_n , t_n), t_n) \rangle_{v^i_L + a \dot n^i_L }  
\ee
so we see that if we choose $a \dot n^i_L = - v^i_L$, we can relate a correlation function with a large scale velocity $v^i_L$ to a correlation function with zero large-scale velocity.  Choosing $n^i_L$ as in \eqn{chooseni} means that we are are considering a physical solution.  Because this relationship does not depend on $\delta^{(1)}$ being a growing mode, we can consider generic Gaussian initial conditions and time dependence
\be
\delta^{(1)}_{\qvec} ( t ) = \left( T_+ ( q) D_+ ( t ) + T_- ( q ) D_- ( t ) \right) \xi_{\qvec} \equiv T(q , t) \xi_{\qvec} \ ,
\ee
where $\xi_{\qvec}$ is a Gaussian field and $T_{\pm} (q)$ are transfer functions for the growing and decaying modes.

In Fourier space \eqn{realcoordchg} becomes
\begin{align}
\begin{split} \label{coorcoortrans}
\langle \delta_{\kvec_1} ( t_1) \cdots  \delta_{\kvec_n} ( t _n) \rangle_{v^i_L} & \approx \left( \prod_{a=1}^n \int d^3 x_a \int_{\kvec_a'} e^{i(\kvec_a' - \kvec_a) \cdot \xvec_a}    \right) \exp \left\{ \sum_{b=1}^n \int_{\qvec} \frac{\qvec \cdot \kvec_b' }{q^2} \delta^{(1)}_{\qvec} ( t_b) e^{i \qvec \cdot \xvec_b} \right\} \\
& \hspace{.5in} \times \langle \delta_{\kvec_1'} ( t_1 ) \cdots \delta_{\kvec_n' } ( t_n) \rangle_0 
\end{split}
\end{align}
To find the single soft-limit consistency relation, we expand the exponential containing $\delta_{\qvec}^{(1)}$ to first order, and to find the double soft-limit relation, we expand it to second order.\footnote{By not expanding the exponential in \eqn{coorcoortrans}, one can keep higher orders in $k \, \delta_{\qvec}^{(1)} / q$ in the consistency relations \cite{Creminelli:2013poa}.}  This gives us the two relevant consistency relations
\be \label{singlecrgd}
\lim_{q_1 \rightarrow 0}  \frac{ \langle \delta_{\qvec_1} ( \tau_1 )  \,  \delta_{\kvec_1 } ( t_1) \cdots \delta_{\kvec_n}(t_n) \rangle '  }{P_{11} ( q_1 ; \tau_1)}  = -  \left(\sum_{a=1}^n  \frac{T(q_1 , t_a)}{T(q_1 , \tau_1) } \frac{ \qvec_1 \cdot \kvec_a}{q_1^2} \right)  \langle   \delta_{\kvec_1 } (t_1) \cdots \delta_{\kvec_n}(t_n) \rangle'_0   \ ,
\ee
and
\begin{align}
\begin{split} \label{doublesoftgd}
& \lim_{q_1,q_2 \rightarrow 0}  \frac{ \langle \delta_{\qvec_1} ( \tau_1 ) \delta_{\qvec_2} ( \tau_2)  \,  \delta_{\kvec_1 } ( t_1) \cdots \delta_{\kvec_n}(t_n) \rangle '  }{P_{11} ( q_1 ; \tau_1) P_{11} (q_2 ;  \tau_2) }  = \\
& \hspace{1.3in}  \left( \sum_{a=1}^n  \frac{T( q_1 , t_a )}{T ( q_1, \tau_1)} \frac{ \qvec_1 \cdot \kvec_a}{q_1^2} \right)   \left( \sum_{b=1}^n \frac{T(q_2,t_b)}{T(q_2,\tau_2)} \frac{ \qvec_2 \cdot \kvec_b}{q_2^2} \right) \langle   \delta_{\kvec_1 } (t_1) \cdots \delta_{\kvec_n}(t_n) \rangle'_0   \ . 
\end{split}
\end{align}

The effect of soft loops on correlation functions \eqn{irloops}, can be computed by averaging \eqn{coorcoortrans} over the long modes
\be
\langle \delta_{\kvec_1} ( t_1 ) \cdots \delta_{\kvec_n} ( t_n ) \rangle_{\text{IR loops}}  \equiv  \langle  \langle \delta ( x^i_1 , t_1) \cdots \delta ( x^i_n , t_n) \rangle_{v^i_L}    \rangle \ . 
\ee
Next, using \eqn{coorcoortrans}, we find 
\begin{align}
\begin{split} \label{irloopsgd}
\langle \delta_{\kvec_1} ( t_1 ) \cdots \delta_{\kvec_n} ( t_n ) \rangle'_{\text{IR loops}}  & \approx \exp \left\{ -\frac{1}{2} \int_{\qvec}\,  \left(\sum_{a=1}^n   T(q, t_a) \frac{\qvec \cdot \kvec_a}{q^2}  \right)^2 \frac{ P_{11}(q; t_{\rm in})}{T(q, t_{\rm in} )^2}  \right\} \\
&\hspace{1in} \times  \langle \delta_{\kvec_1} ( t_1 ) \cdots \delta_{\kvec_n} ( t_n ) \rangle'_0 \ .
\end{split}
\end{align}

%
%
%
\subsection{Perturbative correlation functions in the IR limit}
For illustration, in this section, we use the solutions from \secref{pertirsec} to confirm some examples of the consistency relations perturbatively.  We start with the squeezed limit of the tree-level bispectrum, given in \eqn{btree}.  We have (working in the limit $q , q_1 \ll k_1 , k_2$),
\begin{align}
 \langle \delta_{\qvec} ( \tau ) \delta_{\kvec_1} ( t_1) \delta_{\kvec_2}  ( t_2 ) \rangle'  & \approx  \langle \delta^{(1)}_{\qvec} ( \tau ) \delta^{(2)}_{\kvec_1} ( t_1) \delta^{(1)}_{\kvec_2}  ( t_2 ) \rangle' +  (\{\kvec_1 , t_1 \} \leftrightarrow \{\kvec_2 , t_2 \} )  \\
 & \approx \int_{\qvec_1} \frac{\qvec_1 \cdot  \kvec_1 }{q_1^2} \langle \delta^{(1)}_{\qvec} ( \tau ) \delta^{(1)}_{\qvec_1} ( t_1) \rangle \langle \delta^{(1)}_{\kvec_1 - \qvec_1} ( t_1 ) \delta^{(1)}_{\kvec_2} ( t_2 ) \rangle'  +  (\{\kvec_1 , t_1 \} \leftrightarrow \{\kvec_2 , t_2 \} )  \nonumber  \\
 & \approx - P_{11} ( q ; \tau ) \left(\sum_{a=1}^{2}   \frac{D_+(t_a)}{D_+(\tau)} \frac{\qvec \cdot  \kvec_a}{q^2}  \right)  \langle \delta^{(1)}_{\kvec_1 } ( t_1 ) \delta^{(1)}_{\kvec_2} ( t_2 ) \rangle'  \ , \nonumber
\end{align}
where we have used that $\delta_{\qvec}$ is always in the linear regime, is a growing mode solution \eqn{deltasol}, and that long modes do not correlate with short modes.  This indeed matches the expression given in the non-perturbative consistency relation \eqn{singlecr}.

The calculation for the double soft limit of the tree-level four point function follows in an analogous manner.  This time, the fact that we end up with the form \eqn{doublesoft} is slightly less trivial, since there are two types of diagrams which have to add together with the correct relative coefficient.  We have (working in the limit $q_1, q_2 \ll k_1 , k_2$)
\begin{align} \label{doublesoftpert}
 \langle \delta_{\qvec_1} ( \tau_1 ) \delta_{\qvec_2} ( \tau_2 ) \delta_{\kvec_1} ( t_1) \delta_{\kvec_2}  ( t_2 ) \rangle'   \approx &  \langle \delta^{(1)}_{\qvec_1} ( \tau_1 ) \delta^{(1)}_{\qvec_2} ( \tau_2 ) \delta^{(2)}_{\kvec_1} ( t_1) \delta^{(2)}_{\kvec_2}  ( t_2 ) \rangle'  \\ 
 & + \left(  \langle \delta^{(1)}_{\qvec_1} ( \tau_1 ) \delta^{(1)}_{\qvec_2} ( \tau_2 ) \delta^{(3)}_{\kvec_1} ( t_1) \delta^{(1)}_{\kvec_2}  ( t_2 ) \rangle'  +  (\{\kvec_1 , t_1 \} \leftrightarrow \{\kvec_2 , t_2 \} ) \right) \ .  \nonumber
 \end{align}
 The first diagram is 
 \begin{align}
 \begin{split}
\frac{  \langle \delta^{(1)}_{\qvec_1} ( \tau_1 ) \delta^{(1)}_{\qvec_2} ( \tau_2 ) \delta^{(2)}_{\kvec_1} ( t_1) \delta^{(2)}_{\kvec_2}  ( t_2 ) \rangle' }{P_{11}( q_1 ; \tau_1 ) P_{11} ( q_2 ; \tau_2 )  \langle \delta_{\kvec_1}^{(1)}(t_1)  \delta_{\kvec_2}^{(1)}(t_2) \rangle'  } \approx \frac{D_+ ( t_1 ) D_+ ( t_2 )}{D_+(\tau_1) D_+(\tau_2) } \left( \frac{\qvec_1 \cdot \kvec_1}{q_1^2} \frac{\qvec_2 \cdot \kvec_2}{q_2^2} + \frac{\qvec_1 \cdot \kvec_2 }{q_1^2} \frac{\qvec_2 \cdot \kvec_1}{q_2^2} \right)  \ , 
 \end{split}
 \end{align}
 while the second diagram is 
  \begin{align}
 \begin{split}
\frac{  \langle \delta^{(1)}_{\qvec_1} ( \tau_1 ) \delta^{(1)}_{\qvec_2} ( \tau_2 ) \delta^{(3)}_{\kvec_1} ( t_1) \delta^{(1)}_{\kvec_2}  ( t_2 ) \rangle' }{P_{11}( q_1 ; \tau_1 ) P_{11} ( q_2 ; \tau_2 )  \langle \delta_{\kvec_1}^{(1)}(t_1)  \delta_{\kvec_2}^{(1)}(t_2) \rangle'  } \approx \frac{D_+ ( t_1 )^2}{D_+(\tau_1) D_+(\tau_2) } \frac{\qvec_1 \cdot \kvec_1}{q_1^2} \frac{\qvec_2 \cdot \kvec_1}{q_2^2}  \ .
 \end{split}
 \end{align}
Then, adding all of the pieces in \eqn{doublesoftpert} together gives the final result, which as expected is given by \eqn{doublesoft}.  

As a final example, we look at the one-loop power spectrum, where we can compute the contribution from IR modes in the loop integral.   Using \eqn{p1loopdefs}, we have (working in the limit $q_1, q_2 \ll k_1 , k_2$)
\begin{align}
\begin{split}
\langle \delta^{(2)}_{\kvec_1} ( t_1) \delta^{(2)}_{\kvec_2} (t_2) \rangle'  & \approx \int_{\qvec_1 , \qvec_2} \frac{\qvec_1 \cdot \kvec_1}{q_1^2} \frac{\qvec_2 \cdot \kvec_2}{q_2^2} \langle \delta^{(1)}_{\qvec_1} (t_1) \delta^{(1)}_{\qvec_2} (t_2) \rangle \langle \delta^{(1)}_{\kvec_1 - \qvec_1} ( t_1 ) \delta^{(1)}_{\kvec_2 - \qvec_2} (t_2) \rangle' \\
& \approx  -  \int_{\qvec} \frac{\qvec \cdot \kvec_1}{q^2} \frac{\qvec \cdot \kvec_2}{q^2}  D_+ (t_1) D_+ ( t_2 ) \frac{ P_{11} ( q; t_{\rm in} )}{D_+(t_{\rm in})^2}  P_{11} ( k_1 ; t_1 , t_2)  \ ,
\end{split}
\end{align}
and similarly 
\begin{align}
\begin{split}
 \langle \delta^{(1)}_{\kvec_1} ( t_1 ) \delta^{(3)}_{\kvec_2} (t_2) \rangle'  & \approx   \int_{\qvec_1 , \qvec_2}  \half \frac{\qvec_1 \cdot \kvec_2 }{q_1^2} \frac{\qvec_2 \cdot \kvec_2}{q_2^2} \langle \delta^{(1)}_{\qvec_1} (t_2) \delta^{(1)}_{\qvec_2} (t_2) \rangle \langle \delta^{(1)}_{\kvec_1 } (t_1) \delta^{(1)}_{\kvec_2 - \qvec_1 - \qvec_2}  (t_2) \rangle' \\
& \approx  - \half  \int_{\qvec}  \frac{\qvec \cdot \kvec_2}{q^2} \frac{\qvec \cdot \kvec_2}{q^2} D_+(t_2)^2 \frac{P_{11} ( q; t_{\rm in})}{D_+(t_{\rm in})^2}  P_{11} ( k_1 ; t_1 , t_2)  \ .
\end{split}
\end{align}
Then, adding these to the final contribution $ \langle \delta^{(3)}_{\kvec_1} ( t_1 ) \delta^{(1)}_{\kvec_2} (t_2) \rangle' $ gives exactly \eqn{irloops} expanded to first order in the exponential.

%
%
\section{Green's function manipulations} \label{gfapp}

In this Appendix, we show how to derive the IR solutions for Galilean invariant equations of motion in perturbation theory with exact time dependence presented in \secref{pertirsec}. 
We start with generic linear equations of the form 
\be
\ddot \delta + \bar \nu_\Phi \dot \delta  - \bar \mu_\Phi \delta  = 0 \ , 
\ee
and then introduce the non-linear terms as in \eqn{ireom}, which gives the leading IR equations of motion
\begin{align}
\begin{split} \label{IReom2}
\ddot \delta + \bar \nu_\Phi \dot \delta - \bar \mu_\Phi \delta \approx & - a^{-1} \left( 2 v^i \partial_i \dot \delta + \dot v^i \partial_i \delta - H v^i \partial_i \delta \right) - a^{-2} v^i v^j \partial_i \partial_j \delta  - \bar \nu_\Phi a^{-1} v^i \partial_i \delta  \ . 
\end{split}
\end{align}
Notice that these are the same equations as \eqn{IReom} if the relative velocity is set to zero there.

We will find the solution to \eqn{IReom2} by directly using the Green's function, which satisfies

\be
\partial^2_t G ( t , \ttt ) + \bar \nu_\Phi ( t ) \partial_t G ( t , \ttt ) - \bar \mu_\Phi ( t ) G( t , \ttt ) = \delta_D ( t - \ttt) \ , 
\ee
and has the usual representation in terms of the linear solutions $D_{\pm} ( t ) $ which solve
\be \label{lineards}
 \ddot D_\pm + \bar \nu_\Phi \dot D_\pm - \bar \mu_\Phi  D_\pm = 0 \ . 
\ee
The retarded Green's function is then given by
\be
G(t, \ttt ) = W(\ttt)^{-1} \left( D_- ( t ) D_+ ( \ttt) - D_+ ( t ) D_- ( \ttt) \right) \Theta_{\rm H} ( t - \ttt)  \ ,
\ee
where the Wronskian is given by 
\be
W( t ) = D_+ (t ) \dot D_-(t) - D_-(t) \dot D_+(t) \ .
\ee

As a warm up, let us find the solution for $\delta^{(2)}$.  We have, valid at leading order in $k / q$,
\be
 \delta^{(2)} ( t )  \approx - \int^t d \ttt \, \bar G ( t , \ttt ) \,  \tilde a^{-1} \left(  \dot v^i_{(1)} ( \ttt )\partial_i \delta^{(1)} (\ttt) + 2 v^i_{(1)}(\ttt) \partial_i \dot \delta^{(1)} (\ttt)  + \left(    \bar \nu_\Phi ( \ttt ) -  \tilde H \right) v^i_{(1)} (\ttt) \partial_i \delta^{(1)} (\ttt)  \right)  \ , 
\ee
where we have suppressed the space coordinate $\xvec$, $\tilde a \equiv a ( \ttt)$, and $\tilde H \equiv H ( \ttt)$.  Now we integrate by parts the $\dot v^i_{(1)}$ term to obtain
\be \label{step1}
 \delta^{(2)} ( t )  \approx - \int^t d \ttt  \,  \tilde a^{-1} v^i_{(1)}( \ttt)  \left( ( \bar \nu_\Phi ( \ttt ) \bar G ( t , \ttt)  - \partial_{\ttt} \bar G( t , \ttt) ) \partial_i \delta^{(1)} ( \ttt)  + \bar G ( t , \ttt) \partial_i \dot \delta^{(1)} ( \ttt)   \right)  \ .
\ee
Next, we need an identity for the Green's function.   First of all, using \eqn{lineards} to replace $\ddot D_{\pm}$, we have 
\be
\dot W = - \bar \nu_\Phi  W \ . 
\ee
Using this, we find that
\be \label{magic1}
\left( \bar \nu_\Phi ( \ttt) \bar G ( t , \ttt) - \partial_{\ttt} \bar G( t , \ttt) \right) \delta ( \ttt) + \bar G( t , \ttt ) \dot \delta ( \ttt ) = \delta ( t ) \ 
\ee
for any linear solution $\delta ( \ttt ) = c_+ D_+( \ttt ) + c_- D_-( \ttt) $.   Then, using \eqn{magic1} in \eqn{step1}, we obtain
\be
 \delta^{(2)} ( t )  \approx - \int^t d \ttt  \,  \tilde a^{-1} v^i_{(1)}( \ttt) \partial_i \delta^{(1)} ( t ) 
\ee
and finally using the linear expression for $v^i_{(1)}$ from \eqn{linearvel} and integrating the total derivative, we have 
\be \label{delta2sol}
\delta^{(2)} ( t ) \approx \frac{\partial_i \delta^{(1)} ( t ) }{\partial^2} \partial_i \delta^{(1)} ( t ) \ .
\ee

Now we move on to the general solution.  The formal solution to \eqn{IReom2}, again, valid at leading order in $k / q$, is 
\begin{align}
\begin{split} \label{generalgfinte}
& \delta^{(n+2)} ( t )  \approx - \int^t d \ttt \, \bar G ( t , \ttt ) \,  a( \ttt) ^{-1} \Big\{  \dot v^i_{(1)} \partial_i \delta^{(n+1)} + 2 v^i_{(1)} \partial_i \dot \delta^{(n+1)} + ( \bar \nu_\Phi   - H )  v^i_{(1)} \partial_i \delta^{(n+1)}  \\
& \hspace{4in} + a^{-1} v^i_{(1)} v^j_{(1)} \partial_i \partial_j \delta^{(n)}   \Big\}_{\ttt} 
\end{split}
\end{align}
where we have suppressed the space coordinate $\xvec$, and all time arguments under the integral are evaluated at $\ttt$ unless otherwise shown.\footnote{{Notice that the velocity always appears at first order above.  This is because, in order to find the leading contribution in $k/q$ for a given order in $\delta$, one wants the most number of derivatives on the short mode (to get the most factors of $k$).  As we will show below, the leading solution has the form $\delta^{(n+1)}_{\kvec} \sim ( k / q)^n [\delta^{(1)}_{\qvec}]^{n}\delta^{(1)}_{\kvec}$, and one can explicitly show that any terms with higher than first order in the velocity in \eqn{generalgfinte} lead to subleading contributions.      }}

First, we integrate by parts the $\dot v^i_{(1)}$ term to get
\begin{align}
\begin{split} \label{someeqn}
& \delta^{(n+2) } ( t ) \approx - \int^t d \ttt \, a( \ttt )^{-1} v_{(1)}^i(\ttt) \Big\{ ( \bar \nu_\Phi \bar G( t , \ttt)  - \partial_{\ttt} \bar G( t  , \ttt ) ) \partial_i \delta^{(n+1)}  \\
& \hspace{2in} +\bar G ( t , \ttt) \left( \partial_i \dot \delta^{(n+1)} + a^{-1} v^j_{(1)} \partial_i \partial_j \delta^{(n)} \right)    \Big\}_{\ttt}  \ .
\end{split}
\end{align}
To solve this, we use induction.  Starting with \eqn{delta2sol} as the base case, we assume
\be \label{assume}
\delta^{(n+1)}  \approx \frac{1}{n!} \left( \prod_{a = 1}^n  \frac{\partial_{i_a} \delta^{(1)} }{\partial^2}   \right) \partial_{i_1} \cdots \partial_{i_n}  \delta^{(1)}   \ ,
\ee
for $n>1$, where it is understood that each pair of indices $i_a$ is summed over, and all fields are evaluated at the same time $t$.  

 First consider the last line of \eqn{someeqn}.  Using \eqn{assume}, we have, to leading order in $k / q$,
 \begin{align}
 \begin{split}
 a^{-1} v^j_{(1)} \partial_i \partial_j \delta^{(n)}  & \approx - \frac{\partial_j \dot \delta^{(1)} }{\partial^2} \frac{1}{(n-1)!}  \left( \prod_{a = 1}^{n-1}  \frac{\partial_{i_a} \delta^{(1)} }{\partial^2}   \right) \partial_i \partial_j \partial_{i_1} \cdots \partial_{i_{n-1}} \delta^{(1)} \\
 & = - \frac{1}{n!} \frac{ d}{d t} \left( \prod_{a = 1}^n  \frac{\partial_{i_a} \delta^{(1)} }{\partial^2}   \right) \partial_i \partial_{i_1} \cdots \partial_{i_n}  \delta^{(1)}   \ ,
 \end{split}
 \end{align}
 because of the symmetry of $\partial_i \partial_{i_1} \cdots \partial_{i_n} \delta^{(1)}$.   This means that   
  \be
 \partial_i \dot \delta^{(n+1)}  + a^{-1} v^j_{(1)} \partial_i \partial_j \delta^{(n)} \approx  \frac{1}{n!} \left( \prod_{a = 1}^n  \frac{\partial_{i_a} \delta^{(1)} }{\partial^2}   \right) \partial_i \partial_{i_1} \cdots \partial_{i_n}  \dot  \delta^{(1)}   \ .
  \ee
Then, combining the first and second lines of \eqn{someeqn} and using \eqn{magic1}, we have 
\begin{align}
\begin{split} \label{someeqn2}
& \delta^{(n+2) } ( t ) \approx - \int^t d \ttt \, a^{-1} v_{(1)}^i  ( \ttt )  \frac{1}{n!} \left( \prod_{a = 1}^n  \frac{\partial_{i_a} \delta^{(1)  }  ( \ttt )  }{\partial^2}   \right) \partial_i \partial_{i_1} \cdots \partial_{i_n}    \delta^{(1)} ( t )  \ .
\end{split}
\end{align}
Next, we use the linear solution \eqn{linearvel} for $v^i_{(1)}$ to write 
\begin{align}
\begin{split}
& - a^{-1} v^i_{(1)} ( \ttt)  \frac{1}{n!} \left( \prod_{a = 1}^n  \frac{\partial_{i_a} \delta^{(1)  }  (\ttt)    }{\partial^2}   \right)   \partial_i \partial_{i_1} \cdots \partial_{i_n}    \delta^{(1)} ( t ) \\
&\hspace{1in}  = \frac{1}{(n+1)!} \frac{d}{ d \ttt} \left( \frac{\partial_i \delta^{(1)} (\ttt) }{ \partial^2} \prod_{a = 1}^n  \frac{\partial_{i_a} \delta^{(1)  }  (\ttt)  }{\partial^2}   \right)   \partial_i \partial_{i_1} \cdots \partial_{i_n}    \delta^{(1)} ( t ) \ , 
\end{split}
\end{align}
so that we finally have
\begin{align}
\begin{split}
\delta^{(n+2)} (t ) &  \approx \int^t d \ttt \, \frac{1}{(n+1)!} \frac{d}{ d \ttt} \left( \prod_{a = 1}^{n+1}  \frac{\partial_{i_a} \delta^{(1)  }  ( \ttt )      }{\partial^2}   \right)    \partial_{i_1} \cdots \partial_{i_{n+1}}    \delta^{(1)} ( t )  \\
& =  \frac{1}{(n+1)!}  \left( \prod_{a = 1}^{n+1}  \frac{\partial_{i_a} \delta^{(1)  }  ( t )      }{\partial^2}   \right)    \partial_{i_1} \cdots \partial_{i_{n+1}}    \delta^{(1)} ( t )  \ ,
\end{split}
\end{align}
which was the assumed form, so the induction is complete.

%
%
\section{Coefficients} 
\label{sec:coeff}
Here, we report many of the coefficients appearing in the main text.  We find it useful sometimes to write $\chi_N = \{ \Phi , \Psi \}$ in order to separate the Newtonian potentials from the scalar field $\pi$.  

\subsection{Equations of motion} 
\label{coefficientsapp}
The coefficients appearing in the gravitational equations \eqns{phieq1}{pieq1}, are given explicitly by 
 \begin{align}
c_1 & = -4H \alphaB  + H ( 4 \alphaH - 2 \beta_3 (1 + \alphaM )) - 2 \dot \beta_3 \;, \nonumber   \\
c_2 & = 4 H (1 + \alphaM - c_{\rm T}^2)  +  4 \left(  H    \alphaH (1 + \alphaM)   + \dot \alpha_{\rm H} \right)\;,  \nonumber\\
c_3 & = - 2 H^2 \mathcal{C}_2+ \frac{1}{2} \left\{   H \left[ 4 \dot \alpha_{\rm H} - 2 ( 1 + \alphaM) \dot \beta_3  - \beta_3 \dot {\alpha}_{\rm M} \right] - \ddot \beta_3           \right\}  \\
& \quad \quad + \frac{1}{2} \big\{ - H^2 ( 1 + \alphaM) \left[ - 4 \alphaH + \beta_3 ( 1 + \alphaM)\right]    + 4 \alphaH \dot H - \beta_3 ( 1 + \alphaM) \dot H  \big\} \;,  \nonumber \\
 c_4 & = 4 ( 1 + \alpha_{\rm H} ) \ ,  \qquad  c_5  = -2 c_{\rm T}^2 \ , \qquad  c_6  = - \beta_3 \ ,  \nonumber\\ 
 c_7  &= 4 \alpha_{\rm H} \;, \qquad
 c_8  = -2 (  2 \beta_1 + \beta_3 ) \ , \qquad  c_9  = 4 \beta_1 + \beta_3 \ , \nonumber
 \end{align}
 with
\begin{align}
\mathcal{C}_2 & \equiv  - \alphaM + \alphaB ( 1 + \alphaM ) + c_{\rm T}^2 -1    +  ( 1 + \alphaB)\frac{\dot H}{H^2} + \frac{ \dot \alpha_{\rm B} }{H} + \frac{\bar \rho }{2H^2 M^2}   \ , \\
 b_1 & =  H  \left[ 4 \alphaB + \alphaV (-1+\alphaM)  -2 \alphaM + 3 (c_{\rm T}^2 -1) \right]  + \dot \alpha_{\rm V}   - H \left[  8 \beta_1 \alphaM + \alphaH ( 3 + \alphaM ) \right]  - \dot \alpha_{\rm H} - 8 \dot \beta_1 \;,  \nonumber \\
 b_2 & = \alpha_{\rm V} - \alpha_{\rm H} - 4 \beta_1 \ , \qquad  b_3  = c_{\rm T}^2 -1  \nonumber \ .
\end{align}
We have also defined 
\begin{align}
\begin{split}
C_1 & \equiv  \frac{1}{4} \left( c_1  - H c_8 ( 1 + \alphaM)  - \dot c_8 \right) \;,  \quad  C_2  \equiv \frac{1}{4} \left(   c_2 - H c_7 ( 1 + \alphaM ) - \dot c_7 \right)\;,  \\
C_3 & \equiv  \frac{1}{4} \Big\{ 2 c_3 + (1 + \alphaM) \left[ 2 H \dot c_9 + c_9 \left( H^2 (1 + \alphaM)  + \dot H \right)  \right]  + c_9 H \dot \alpha_{\rm M} + \ddot c_9     \Big\}  \;,  \\
C_4 & \equiv \half ( c_9 H ( 1 + \alphaM) + \dot c_9 ) \;.
\end{split}
\end{align}
Note that these equations are general: no degeneracy conditions or observational constraints have been assumed.

%
%
\subsection{Linear solutions} \label{linearsolssec}

Here, we focus on the linear equations of motion to give explicit expressions for the coefficients appearing in \eqn{linearsols}.  As usual, we solve \eqns{phieq1}{psieq1} for $\partial^2 \Phi$ and $\partial^2 \Psi$ in terms of $\partial^2 \pi$, $\partial^2 \dot \pi$, and $\delta$.  This has the form
\begin{align}
\begin{split}
\partial^2 \chi_N = \omega^{ \chi_N}_{1} \partial^2 \pi + \omega^{ \chi_N}_2 \partial^2 \dot \pi + \omega^{ \chi_N}_3 \delta  \ , 
\end{split}
\end{align}
where 
\begin{align}
\begin{split}
& \omega^\Phi_1 = \frac{ 8 C_1 c_5 -4 C_2 c_4 }{\omega} \ , \quad \omega^\Phi_2 = \frac{c_4 c_7 - 2 c_5 c_8 }{\omega} \ , \quad  \omega^\Phi_3 = \frac{-4 a^2 c_5 \barrhom}{\omega M^2} \ , \\
& \omega^\Psi_1 = \frac{ 8 C_2 c_6  -4 C_1 c_4 }{\omega} \ , \quad  \omega^\Psi_2 = \frac{c_4 c_8 - 2 c_6 c_7 }{\omega} \ , \quad \omega^\Psi_3 = \frac{2 a^2 c_4 \barrhom }{\omega M^2} \ ,
\end{split}
\end{align}
where $\omega \equiv c_4^2 - 4 c_5 c_6$.  
Next, we plug these expressions into \eqn{pieq1} to obtain the expression for $\partial^2 \pi$, where, once we impose the degeneracy conditions discussed in \secref{introsec}, the terms proportional to $\partial^2 \dot \pi$ and $\partial^2 \ddot \pi$ drop out, and we are left with an expression as in \eqn{linearsols} with
\begin{align}
\begin{split}
\mu_\pi &  =-  \frac{c_1\omega^\Phi_3 + c_2 \omega^\Psi_3 + c_8 \dot \omega^\Phi_3 + c_7 \dot \omega^\Psi_3}{ a^2 C_\pi} \ , \quad \text{and} \quad  \nu_\pi   = -  \frac{c_8 \omega^\Phi_3 + c_7 \omega^\Psi_3 }{a^2 C_\pi} \ , 
\end{split}
\end{align}
where we have defined
\be
C_\pi = 4 C_3 + c_1 \omega^\Phi_1 + c_2 \omega^\Psi_1 + c_8 \dot \omega^\Phi_1 + c_7 \dot \omega^\Psi_1 \ . 
\ee

 Next, we plug the solution for $\partial^2 \pi$ into \eqns{phieq1}{psieq1} to get the solutions for $\partial^2 \Phi$ and $\partial^2 \Psi$ in the form \eqn{linearsols} with
 \begin{align}
 \begin{split}
 \mu_{\chi_N} &=  \mu_\pi (\omega^{\chi_N}_1 + 2 H \omega^{\chi_N}_2 ) + \dot \mu_\pi  \omega^{\chi_N}_2+ a^{-2}\omega^{\chi_N}_3  \ , \\
 \nu_{\chi_N} &= \mu_\pi \omega^{\chi_N}_2 + \nu_\pi (\omega^{\chi_N}_1 + 2 H \omega^{\chi_N}_2 ) + \dot \nu_\pi \omega^{\chi_N}_2 \ , \\
 \sigma_{\chi_N} & = \nu_\pi \omega^{\chi_N}_2 \ . 
 \end{split}
 \end{align}
Note that these equations are general: no degeneracy conditions or observational constraints have been assumed, except to say that the terms proportional to $\partial^2 \dot \pi$ and $\partial^2 \ddot \pi$ drop out of the solution for $\partial^2 \pi$.  

Now, as an example, we specialize to the case where we impose the degeneracy conditions discussed in the Introduction, along with $\alphaH = - 2 \beta_1$ (which imposes that gravitational waves do not decay).  This gives
\begin{align}
\mu_\pi & =  \frac{1}{M^2 C_\pi (1 - \beta_1)^2} \Big(  2(1- \beta_1)\beta_1 \dot{\bar \rho} \nonumber  + 2 H \barrhom  (\alphaB - \alphaM ( 1 - \beta_1) + \beta_1 (4 - 3 \beta_1))    \Big)  \nonumber \\
\nu_\pi & = \frac{2 \beta_1 \barrhom}{M^2 C_\pi ( 1 - \beta_1 ) } \
\end{align}
where
\be
C_\pi = \frac{2 H^2 \alpha c_s^2}{(1-\beta_1)^2 } \ , 
\ee
and
\begin{align}
 \alpha  c_s^2   \equiv - \frac{\bar \rho_{\rm m} ( 1 - \beta_1)^2}{ H^2 M^2}  + 2   \left( 1 + \alphaB - \frac{\dot \beta_1 }{H} \right)^{2}   \bigg[  \frac{1}{a M^2} \frac{d}{dt }   \left( \frac{a M^2 (1-\beta_1)}{H ( 1 + \alphaB) - \dot \beta_1}  \right)  -1  \bigg]  \nonumber  \ . 
\end{align}

For the other coefficients in \eqn{linearsols}, we have 
\begin{align}
\begin{split}
\mu_{\Phi} &  = \frac{\barrhom}{2 M^2 (1-\beta_1)^2} + \frac{\mu_\pi \varpi_\Phi - \dot \mu_\pi \beta_1 }{1 - \beta_1 } \ , \\
\nu_\Phi & = \frac{ - ( \mu_\pi + \dot \nu_ \pi )\beta_1 + \nu_\pi \varpi_\Phi}{1-\beta_1} \ ,  \,\,\,\, \sigma_\Phi  = - \frac{\nu_\pi \beta_1}{1-\beta_1} \ , 
\end{split}
\end{align}
and 
\begin{align}
\begin{split}
\mu_{\Psi} &  = \frac{\barrhom(1-2 \beta_1)}{2 M^2 (1-\beta_1)^2} + \frac{\mu_\pi \varpi_\Psi + \dot \mu_\pi \beta_1 }{1 - \beta_1 } \ , \\
\nu_\Psi & = \frac{  ( \mu_\pi + \dot \nu_ \pi )\beta_1 + \nu_\pi \varpi_\Psi}{1-\beta_1} \ ,  \quad \sigma_\Psi  =  \frac{\nu_\pi \beta_1}{1-\beta_1} \ , 
\end{split}
\end{align}
where we have defined 
\begin{align}
\begin{split}
\varpi_\Phi & = \frac{H(\alphaB - \alphaM -\beta_1 (1 -\alphaM  - 2 \beta_1)) - \dot \beta_1 }{1-\beta_1} \ ,  \\
\varpi_\Psi & = \frac{1}{1-\beta_1} \Big(  - \dot \beta_1 + 2 \beta_1 \dot \beta_1  + H(\alphaB +\beta_1 (3-2\alphaB +\alphaM ) - \beta_1^2(4 + \alphaM) ) \Big) \ . 
\end{split}
\end{align}

%
\section{Details for IR-resummation} \label{resumapp}

Here we collect some results leading to the power spectrum with the long-wavelength dark-matter velocity resummed in \eqn{irresumnew}.  The first term in  \eqn{newresumexpand1} is the same as the $\Lambda$CDM case, since $\Delta \tilde \delta_{\kvec}$ does not depend on the long modes, so we have
\be
\left\langle E[\delta^{(1)}  ] \, \Delta \tilde \delta_{\kvec_1 '} \Delta \tilde  \delta_{\kvec_2'} \right\rangle    {\approx} \left\langle E[\delta^{(1)} ] \right\rangle \left\langle \Delta \tilde \delta_{\kvec_1 '}  \Delta \tilde  \delta_{\kvec_2'} \right\rangle   \ .
\ee
This will lead to a contribution analogous to the $\Lambda$CDM case given in \eqn{lcdmresum}.  

In the second line of \eqn{newresumexpand1}, however, there is a long mode in $\tilde \delta^{(2)}_{{\rm IR},\kvec'}$ that correlates with $E[\delta^{(1)}  ] $, and in the third line of \eqn{newresumexpand1} there are two long modes that correlate with $E[\delta^{(1)}  ] $.  Thus, overall, we have to compute
\begin{align} \label{termstocompute}
\left\langle E[\delta^{(1)}  ] \right\rangle \ , \quad \left\langle E[\delta^{(1)}  ] \,\delta^{(1)}_{\qvec} \right\rangle \ , \quad \text{and} \quad  \left\langle E[\delta^{(1)}  ] \,\delta^{(1)}_{\qvec_1} \delta^{(1)}_{\qvec_2} \right\rangle \ . 
\end{align}
Terms of this form can be straightforwardly computed using the cumulant theorem and relevant functional derivatives \cite{Creminelli:2013poa}, which we review here.

First, it is convenient to define
\be
J(\qvec  ) \equiv \sum_{b=1}^2 \frac{\qvec \cdot \kvec_b'}{q^2} e^{i \qvec \cdot \xvec_b} \ , \quad \text{so that} \quad
E[ \delta^{(1)} ] = \exp \left\{ \int_{\qvec} J(\qvec) \delta^{(1)}_{\qvec} \right\} \ .
\ee
Then using the cumulant theorem, we have
\be
\langle E[ \delta^{(1)} ]  \rangle = \exp \left\{ \half \int_{\qvec} J( \qvec ) J ( - \qvec ) P_{11} ( q ) \right\} \ .
\ee
Next, we have
\be
\quad \left\langle E[\delta^{(1)}  ] \,\delta^{(1)}_{\qvec} \right\rangle  = ( 2 \pi)^3 \frac{\delta \langle  E[ \delta^{(1)} ]  \rangle }{\delta J(\qvec) } = J(- \qvec) P_{11} ( q) \langle  E[ \delta^{(1)} ]  \rangle \ , 
\ee
and
\begin{align}
\quad \left\langle E[\delta^{(1)}  ] \,\delta^{(1)}_{\qvec_1}  \delta^{(1)}_{\qvec_2} \right\rangle  & = ( 2 \pi)^6 \frac{\delta^2 \langle  E[ \delta^{(1)} ]  \rangle }{\delta J(\qvec_1)  \delta J( \qvec_2)  } \\
& = (2 \pi)^3 \delta_D ( \qvec_1 + \qvec_2 ) P_{11}(q_1) \langle  E[ \delta^{(1)} ]  \rangle  + J(-\qvec_1) J(-\qvec_2) P_{11}(q_1) P_{11}(q_2) \langle  E[ \delta^{(1)} ]  \rangle  \ . \nonumber
\end{align}

Now, since in the computation \eqn{resum1} we always work to leading order in $k/q$, we can set $\kvec_1' \approx - \kvec_2'$ in the factors of $J(\qvec)$.  This gives
\begin{align}
\begin{split}
\langle  E[ \delta^{(1)} ]  \rangle & \approx \exp \left\{ - \half \int_{\qvec} P_{11}(q) \left( \frac{\qvec \cdot \kvec'_1}{q^2} \right)^2 \left( 2 - e^{i \qvec \cdot (\xvec_1 - \xvec_2 ) } - e^{-i \qvec \cdot (\xvec_1 - \xvec_2) } \right) \right\} \ , \\
\quad \left\langle E[\delta^{(1)}  ] \,\delta^{(1)}_{\qvec} \right\rangle & \approx - \frac{\qvec \cdot \kvec_1'}{q^2} \left( e^{-i \qvec \cdot \xvec_1} - e^{- i \qvec \cdot \xvec_2} \right) \langle  E[ \delta^{(1)} ]  \rangle  \ ,\\
\quad \left\langle E[\delta^{(1)}  ] \,\delta^{(1)}_{\qvec_1}  \delta^{(1)}_{\qvec_2} \right\rangle & \approx  (2 \pi)^3 \delta_D ( \qvec_1 + \qvec_2 ) P_{11}(q_1) \langle  E[ \delta^{(1)} ]  \rangle  \\
& + \left( \frac{\qvec_1 \cdot \kvec_1'}{q_1^2} \frac{\qvec_2 \cdot \kvec_1'}{q_2^2} \right) \left( e^{-i \qvec_1 \cdot \xvec_1} - e^{- i \qvec_1 \cdot \xvec_2} \right)  \left( e^{-i \qvec_2 \cdot \xvec_1} - e^{- i \qvec_2 \cdot \xvec_2} \right) \\
& \hspace{2in} \times P_{11}(q_1) P_{11} ( q_2) \langle  E[ \delta^{(1)} ]  \rangle \ . 
\end{split}
\end{align}


 \bibliographystyle{utphys_UPDATE}
 \small
\bibliography{matt_master_bib}

\end{document}